\documentclass{emulateapj}
\usepackage{natbib}

\def\kms{km s$^{-1}$}
\def\ergs{erg s$^{-1}$}
\def\reff{R$_{eff}$}
\def \mass {M$_*$}
\def \cmd {CMD$_{*}$}
\def\dcol {$\Delta$U-B}
\def\lhard {L$_{2\textrm{-}10}$}

\slugcomment{To appear in Astrophysical Journal}

\shorttitle{CANDELS: AGN Host Properties}

\begin{document}

\title{X-ray selected AGN Hosts are Similar to Inactive Galaxies out to $z=3$: Results from CANDELS/CDF-S}

\author{D.J. Rosario\altaffilmark{1}, M. Mozena\altaffilmark{2}, S. Wuyts\altaffilmark{1}, K. Nandra\altaffilmark{1}, A. Koekemoer\altaffilmark{3}, E. McGrath\altaffilmark{2}, N.P. Hathi\altaffilmark{4}, 
A. Dekel\altaffilmark{5}, J. Donley\altaffilmark{3}, J.S. Dunlop\altaffilmark{6}, S.M. Faber\altaffilmark{2},  H. Ferguson\altaffilmark{3}, M. Giavalisco\altaffilmark{7}, N. Grogin\altaffilmark{3}, Y. Guo\altaffilmark{7}, 
D.D. Kocevski\altaffilmark{2}, D.C. Koo\altaffilmark{2}, E. Laird\altaffilmark{9}, J. Newman\altaffilmark{8}, 
C. Rangel\altaffilmark{9}, R. Somerville\altaffilmark{3}}

\altaffiltext{1}{Max-Planck-Institute for Extraterrestrial Physics,
Garching, 85748}

\altaffiltext{2}{Astronomy Department and UCO-Lick Observatory, University of California, Santa Cruz, CA 95064}

\altaffiltext{3}{Space Telescope Science Institute, USA}

\altaffiltext{4}{Observatories of the Carnegie Institution of Washington, USA}
\altaffiltext{5}{The Hebrew University, Israel}
\altaffiltext{6}{University of Edinburgh, UK}
\altaffiltext{7}{University of Massachusetts, USA}
\altaffiltext{8}{University of Pittsburgh, USA}
\altaffiltext{9}{Imperial College, London, UK}

\begin{abstract}
We use multi-band spatially resolved photometry from the Cosmic Assembly Near-IR Deep Legacy Survey (CANDELS)
in the 4 MSec Chandra Deep Field-South (CDF-S) to explore the nuclear and
extended colors, color gradients and stellar populations of X-ray selected AGN host galaxies out to z=3. 
Based on a study of their central light, we develop X-ray based criteria to exclude objects with strong AGN
contamination. We use stellar masses from the FIREWORKS database to 
understand and account for stellar mass selection effects, and carefully study, for the first time, 
the resolved host galaxy properties of AGNs at $z\sim2$ in their rest-frame optical light
without substantial nuclear contamination. AGN hosts span a sizable range of stellar masses,
colors and color gradients at these redshifts. Their colors, color gradients and stellar population properties are 
very similar to inactive galaxies of the same stellar mass. At $z\sim1$, we find a slightly narrower range in
host colors compared to inactive galaxies, as well as hints of more recent star-formation. These differences
are weaker or non-existent among AGN hosts at $z\sim2$. 
We discuss the importance of AGN driven feedback
in the quenching of galaxies at $z\gtrsim1$ and speculate on possible evolution in the relationship between black
hole accretion and the host galaxy towards high redshifts.

\end{abstract}


\section{Introduction}

Active Galactic Nuclei (AGNs) are the result of accretion onto super-massive black holes (SMBHs), which are generally found in the centers of massive galaxies. 
Till quite recently, activity in galactic nuclei was studied as a topic of special interest, usually unrelated to the evolution of galaxies. 
This has changed in the last decade has changed as tight scaling relationships were uncovered between the masses of SMBHs and the masses of their host spheroids \citep[e.g.][]{magorrian98, ferrarese00, gebhardt00}. These relationships seem to imply a close conection between the evolution of nuclear black holes and the evolution of galaxies as a whole \citep[but, see][]{jahnke11}. Such a connection is, at first glance, quite remarkable, since the typical sphere of influence of even the largest black holes is less than a few tens of parsecs in size, more than 3 orders of magnitude smaller than the typical sizes of galaxies. However, since the inflow of gas to the black hole is the key driver of SMBH growth and this depends on the circum-nuclear environment and the large-scale properties of the host galaxy (e.g., its gas content), an essential connection between black hole growth and host galaxy properties may arise through the fueling of SMBHs.

Another avenue through which nuclear activity can influence the host galaxy, despite the large difference in spatial scales, is through a set of physical processes collectively called `feedback'. Examples include energy and momentum driven outflows in QSOs \citep{pounds03}, outflows accelerated by relativistic jets \citep{morganti05, rosario10} and the suppression of cluster cooling flows by radio lobes \citep{mcnamara07}. 

AGN feedback can play an important role in galaxy evolution by driving and regulating the transformations of star-forming galaxies into quiescent galaxies. Strong feedback from powerful AGNs can eject gas from their host galaxies, effectively shutting down star-formation and moving them onto the Red Sequence \citep{schawinski07, hopkins08b, kaviraj11}. In addition, feedback can prevent star formation by keeping enough gas from cooling and accreting onto these galaxies, thereby keeping them on the Red Sequence. AGN feedback has become an essential element of several modern semi-analytic models of galaxy formation \citep{bower06, croton06, somerville08, cattaneo09}, since it curtails the formation of very high-mass blue galaxies, as required by observations \citep{benson03}. The similarity between the time-scale estimates of AGN activity and the quenching of star-formation also lends some credence to the relevance of AGN feedback \citep{bundy08}.


Several studies have searched for empirical signature of AGN feedback, though, there is a lack of sufficient evidence for 
wide-spread AGN-driven outflows at the level needed to satisfy the requirements of most semi-analytic models \citep{tadhunter08}, 
except in perhaps the most powerful AGNs and in Ultra-luminous Infrared galaxies \citep{rupke11, sturm11}. This may be because evolutionary models invoke a feedback prescription that is too strong, or it may be because outflows are rather transient events and difficult to characterize. An alternate approach has been to search for a direct link between the strength of nuclear activity and the transformation of the galaxy on larger scales by the examination of the structure and star-formation histories of AGN host galaxies. While indirect, these methods are more general and can be applied to large galaxy samples across a range of redshifts.

In the local Universe, AGNs tend to be preferentially in galaxies that lie between the two peaks of the bi-modal galaxy color distribution, i.e, they tend to hosted by so-called `Green Valley'  galaxies \citep{kauffmann03a, schawinski09}. This trend is found among X-ray selected AGNs even out to $z\sim2$ \citep{silverman08, nandra07, brusa09}. Galaxies in the Green Valley are generally believed to be transitioning between from a state of on-going star-formation towards quiescence \citep[but see][]{silverman08}. Therefore, the over-representation of AGNs in the Green Valley is taken as evidence of the influence of the AGN in the suppression of star-formation, though some doubts remain about the generality of this interpretation \citep{xue10, cardamone10}. 

Morphologically, local low-luminosity AGN 
are found in disk galaxies with substantial bulge components, i.e, early-type disks \citep{whittle92, hunt99, schawinski10}. There is, however, considerable scatter in the distribution of AGN host morphologies:  a large proportion, as much as 30\%, have late-type spiral morphologies. Such trends are found among AGN hosts even out to $z\sim2$ \citep{pierce07, georgakakis09, gabor09,schawinski11,kocevski12} and suggest that low and intermediate luminosity AGN are not intimately associated with major galaxy mergers, and that secular processes may play an important role in their fueling. The case is not as clear for luminous AGNs, which are rare, typically at higher redshifts and in which contamination from a nuclear point source can complicate the measurement of host galaxy structure. A few careful studies of the hosts of such AGNs 
suggest a similar distribution of morphological types \citep{dunlop03, guyon06}, though with more pronounced signatures of recent 
galaxy mergers or interactions \citep{bennert08}. 


Evolutionary studies show that the space density of luminous AGN peaks at $z\sim 1.5-2.5$ \citep[e.g.][]{boyle98, hopkins07}. It is at these redshifts that the most pronounced and widespread signatures of AGN feedback on galaxy populations may be expected. Till recently, detailed studies of AGN host galaxies at these epochs have been hampered by their faintness and and the paucity of adequate samples with good redshifts. In addition, at $z>1.5$, the 4000 \AA\ break, a principal spectral diagnostic feature in the stellar continuum of galaxies, is redward of 1 $\mu$. 
Imaging of these distant galaxies with high-resolution instruments, such as those on the Hubble Space Telescope (HST), were generally restricted to the optical bands, which traced the rest-frame near-UV and were most sensitive to the emission from massive stars and recent star-formation. Our understanding of the extended stellar populations of AGN hosts at $z\sim 2$ was restricted to that of a handful of objects imaged with the second-generation HST/NICMOS camera or through Adaptive Optics instruments on large ground-based facilities.

The near-IR channel of the recently commissioned Wide-Field Camera 3 (WFC3) on the refurbished HST is greatly enchancing our view of the distant Universe by providing unprecedented sensitivity and resolution out to 1.7 $\mu$m. 
In this study, we combine the Chandra Deep Field South (CDF-S) 4 MSec X-ray catalog with multi-band optical and NIR imaging from HST Advanced Camera for Surveys (ACS) and WFC3, as well as existing multiwavelength ground-based datasets in the GOODS-S fields. We explore resolved photometry of AGN hosts in the redshift range of $0.5 < z < 3$ and study the properties of their galaxy light without substantial contamination from nuclear AGN emission. We derive rest-frame UV-optical colors and extinctions for these galaxies, constrain their star-formation histories (SFHs) and compare these properties to a well-defined comparison sample of inactive galaxies to place the AGN hosts in the context of field galaxy samples over the range of epochs in which most of the growth of SMBHs and stellar mass occurs. 

The paper is organized as follows: in Section 2, we introduce the various datasets that we bring to bear in this study; in Section 3, we discuss the selection of an AGN sample and its X-ray properties, as well as process of defining a control sample of inactive galaxies; in Section 4 and 5, we introduce the technique of aperture photometry applied to the resolved images of galaxies and the method used to derive various properties such as colors and SFHs. In Section 6, we introduce nuclear colors and define a method to remove objects where AGN light strongly contaminates the extended photometry. Finally, in Section 7, 8 and 9, we analyze the outer colors, extinctions, color gradients and star-formation histories of AGN hosts. We discuss our results in Section 10.

Throughout this work, we adopt a $\Lambda$CDM Concordance cosmology with H$_0 = 70$ \kms Mpc$^{-1}$.


\section{Datasets}

\subsection{HST imaging}

The Wide-Field Camera 3 Infra-Red Channel (WFC3-IR) is a fourth generation instrument
on the HST, designed to provide low thermal background diffraction-limited imaging
over a relatively wide area ($136'' \times 123''$) in the near-IR from 0.8--1.7 $\mu$m.
Details of the instrument and its capabilities can be found in \citet{baggett08}
and the \anchor{http://www.stsci.edu/hst/wfc3/documents/handbooks/currentIHB/wfc3_cover.html}
{WFC3 Instrument Handbook}. 

As part of the CANDELS Multi-Cycle Treasury Survey \citep{grogin11, koekemoer11}, the
GOODS-S field was imaged with two NIR filters - F125W (approx. J band) and F160W (approx. H-band). The
imaging dataset consists of a core set of deep exposures (part of CANDELS-Deep) and a shallower extension
(CANDELS-Wide). Combined with the archival WFC3 Early Release Science (ERS2) dataset \citep{windhorst11}, the
J and H-band imaging covers, in total, 176 arcmin$^2$. The CANDELS-Deep and ERS2 imaging has an average
H band integration of 5 ksec, while the CANDELS-Wide extension has a depth of 1.4 ksec. 
The J band images have a similar depth. The individual WFC3-IR exposures were registered,
cleaned and combined with a custom-built MULTIDRIZZLE-based pipeline into a single large mosaic of the GOODS-S field. 
Details of the survey design, depths, dither patterns, reduction, astrometric and photometric calibration, and mosaic creation can be found in the 
principal CANDELS reference publications \citep{grogin11, koekemoer11}. For this work, we employed 
the combined mulit-depth mosaic and its associated weightmap. 
In addition to CANDELS WFC3 data, we also use imaging in the short Y-band filter (F098M) from the archival ERS2 program.


We also made use of the GOODS ACS v2.0 imaging dataset, publicly
available from the the online \anchor{http://archive.stsci.edu/prepds/goods/}{GOODS database}. 
Images in 4 optical filters (F435W, F606W, F775W and F850LP) were used, with average 
integrated exposures of 7.2, 5.4, 7.0 and 18.2 ksec respectively. Descriptions
of the dataset, including reductions and calibrations, can be found on the public GOODS archive
and in \citet{giavalisco04}. 

The multi-epoch nature of the CANDELS dataset also produces a non-uniform Point Spread Function (PSF) across the CANDELS mosaic,
as subexposures of different depth and orientation are combined at any given pixel location. Detailed modeling of stars 
indicate that the WFC3 PSF is quite stable across each of the subfields. This is because the WFC3 images for each epoch were
generated from tiling patterns with little overlap, ensuring that submosaics from each epoch have uniform depths and most
differences between the PSFs in each epoch were a consequence of the changing orientations \citep{grogin11}. 
On the other hand, the PSFs in each of the three subfields differ greatly between themselves. 
Therefore, we relied on a set of WFC3 PSFs derived for each of the three subfields, one for each band. These representative 
PSFs were constructed from model PSFs for the center of the WFC3-IR camera, generated by the TinyTim software \citep{hook08},
which were then drizzled together using the appropriate dither pattern for the CANDELS submosaics \citep{koekemoer11}.
Appropriate PSFs for the GOODS ACS mosaics were created in the same way as the WFC3 mosaics.
 In \S4, we describe how the PSFs were used to generate kernels to match the resolutions of the HST images in the different bands.



\subsection{FIREWORKS photometric database}

FIREWORKS is a $K_S$-band selected multiwavelength photometric catalog for the CDF-S region,
with matched aperture photometry from imaging in the WFI $U_{38}BVRI$, ACS $B_{435}V_{606}i_{775}z_{850}$,
ISAAC $JHK_S$, all four IRAC channels and MIPS 24$\mu$m bands \citep{wuyts08}. The catalog depth
in the  $K_S$-band is 24.3 mag (AB) and it covers an area of 138 arcmin$^2$, overlapping
considerably, but not completely, with the CANDELS GOODS-S footprint. 
The eastern and western edges of the ERS2 strip and a sizable wedge in the NE corner are not included in FIREWORKS, primarily
because the WFC3 coverage is wider, in the EW direction, than both the GOODS-S ACS
and ISAAC NIR imaging.

For the the full multiwavelength catalog, as well as details of the PSF matching, 
astrometric registration and photometry used to construct the catalog, we refer the reader to \citet{wuyts08}.
The FIREWORKS catalog includes photometric redshift estimates for all sources, estimated
with the EAZY software package \citep{brammer08}. The catalog is supplemented
with spectroscopic redshifts from numerous efforts in the GOODS-S field. The photometric redshifts
are highly accurate, with a median offset in $(z_{spec} - z_{phot})/(1+z_{spec})$ of 0.001, with 
an rms scatter of 0.032, increasing to 0.05 at $z\sim3$. Only a few percent of objects have
catastrophic redshift errors. In this work, we employ both photometric and spectroscopic redshifts
from the FIREWORKS database, using the former only when the latter is unavailable. 
For AGNs, we also included spectroscopic redshifts from more recent campaigns available in the literature
\citep{silverman10}.
In addition to redshifts, we adopt stellar mass estimates for galaxies based on 
FIREWORKS photometry and redshifts. 


\subsection{Chandra 4 MSec X-ray catalog}

The CDF-S 4 MSec dataset consists of 54 separate exposures from the Advanced CCD Imaging Spectrometer imaging array (ACIS-I) taken between October 1999 and July 2010 as an archival resource for the community. The imaging covers an area of 464 arcmin$^2$ centered on $\alpha_{J2000} = 03:32:28.1$, $\delta_{J2000} = -27:48:26.0$, overlapping with the GOODS-S survey field. We reduced the CDF-S archival dataset and extracted a point source catalog from combined image. The methods used in the reduction and creation of the catalog, including the determination of PSFs, source detection and photometry, mirror those developed for the AEGIS-X survey \citep{laird09}. The catalog consists of 571 sources with on-axis limiting depths of $10^{-17}$ \ergs\ in the soft (0.5-2 keV) band and $1.9\times10^{-17}$ erg/s in the hard (0.5-2 keV) band. 

\section{X-ray sources in CANDELS/GOODS-S field}

\begin{figure}[t]
\figurenum{1}
\label{kmag_vs_z}
\centering
\includegraphics[width=\columnwidth]{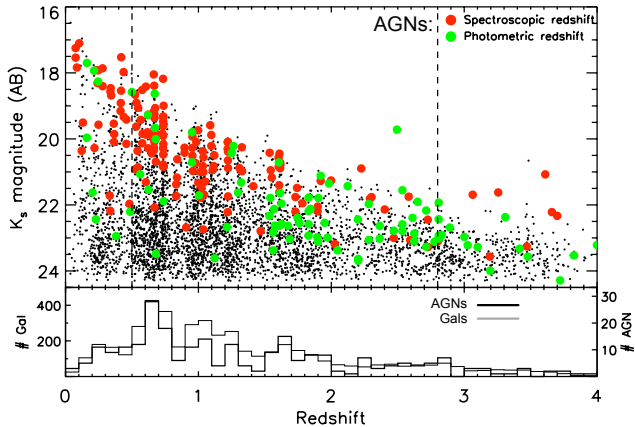}
\caption[Kmag vs. redshift]
{ Apparent K$_s$ magnitude plotted against redshift for X-ray sources detected in the CDF-S 4 Msec dataset (large points),
as well as normal X-ray undetected galaxies from the FIREWORKS catalog (small points). The red points are X-ray sources
with a counterpart with a spectroscopic redshift, while the green points are sources with a photometric redshift.
The histograms in the lower panel plots the redshift distribution of the FIREWORKS galaxies (thin line, left Y-axis) and the AGNs
(thick line, right Y-axis). The dashed lines mark the redshift range to which we restrict this study ($0.5<z<2.8$).
}
\end{figure}

\subsection{Near-IR counterparts of the X-ray sources}

The ACIS-I PSF varies with location on the detector and can be as large as 8" (50\%\ encircled
energy diameter) at large off-axis angles, which reduces both the spatial resolution and sensitivity to point 
sources for sources near the edges of the detector. This axis dependent resolution has to be taken into account 
when matching X-ray sources to near-IR counterparts in the FIREWORKS
catalog. To do this, we adopt a Bayesian cross-matching algorithm, which takes into account the 
positional uncertainty of the X-ray source (which is a function of the X-ray PSF and the presence of
any nearby sources), the offset of possible counterparts as well as the likelihood of a
chance alignment, based on the mean number densities of sources at the K-band magnitudes of the possible counterparts. 
We also visually examined and verified all the cross-matches on the CANDELS H-band images, as well as removed any objects
with bad imaging data.

Of the 571 X-ray sources in the 4 MSec point source catalog, 304 lie in the
area of the field overlapped by the FIREWORKS catalog, the GOODS-S and CANDELS imaging. 
253 sources (83\%) are reliably matched to FIREWORKS counterparts highlighting the efficiency at which NIR imaging yields good
counterparts to X-ray sources. This is because a substantial fraction of sources in deep X-ray catalogs 
are at redshifts greater than 1.5, which makes them increasing faint to optical imaging surveys, but are easily detectable
in near-IR bands. Of the remaining 51 sources not matched to FIREWORKS counterparts, a major fraction appear to be associated with
large bright galaxies or bright stars. 
A small fraction appear to be associated with faint sources in the H-band images which have no FIREWORKS counterparts. 
These are good candidates for high redshift X-ray sources, but are beyond the scope of this work.

For the AGNs, standard photometric redshift techniques, such as those in EAZY, can yield results that are
systematically in error \citep{salvato09, luo10}, due to contamination from the active nucleus.
In the CDF-S, \citet{luo10} determine redshifts for 2 Msec X-ray point sources using a set of templates specially
designed for the AGN population. We compared redshifts for sources from \citet{luo10} with matching FIREWORKS
counterparts to 4 MSec sources. For objects where the redshifts differed by more than 0.05, we adopted the \citet{luo10}
redshifts and recalculated stellar masses using the FIREWORKS machinery. 

In Fig.~\ref{kmag_vs_z}, we plot the FIREWORKS $K_{s}$ magnitude of the cross-matched X-ray sources as a function of 
redshift and compare them to the distribution of all galaxies in the FIREWORKS catalog. 
X-ray sources tend to be among the brightest objects
at any given redshift. This implies that  X-ray AGN are generally found in hosts that are luminous and relatively massive, 
as shown in many previous studies \citep[e.g,][]{nandra07, brusa09}. 

At $z<1.5$, the majority of X-ray source counterparts have redshifts that are determined 
through spectroscopic methods and are highly reliable. An increasingly 
larger fraction of sources at $z>1.5$ have photometric redshifts. 
Given the low error and failure rate of FIREWORKS photometric redshifts, this fact
should not introduce any significant systematic biases in the SED fits or color trends
(we test this in \S6.2). 
For our purposes, we treat both spectroscopic and photometric redshifts equivalently in the
following discussions. 
 
For this study, we restrict ourselves to galaxies with redshifts between 0.5 and 2.8, indicated in Fig.~\ref{kmag_vs_z}
with dashed vertical lines. The low end of this range marks the redshift at which a substantial fraction 
of X-ray emission may arise 
from star-formation or weak AGN activity (L$_{{2\textrm{-}10}} < 10^{42}$ \ergs). The
upper end is the redshift at which the rest-frame 4000 \AA\ break enters the observed H-band (the reddest
band we consider for aperture photometry). 
By applying an upper redshift cut, we restrict our sample to redshifts at which the 
H-band traces light from the bulk of a galaxy's stars.
Our redshifts cuts 
yields a working sample of 176 X-ray sources. 

\begin{figure}[t]
\figurenum{2}
\label{lx_vs_z}
\centering
\includegraphics[width=\columnwidth]{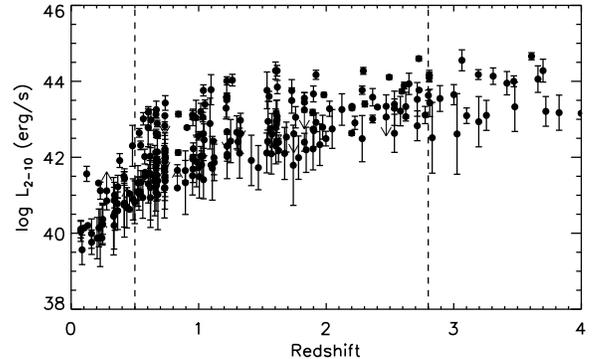}
\caption[L_x vs. redshift]
{ Rest-frame X-ray luminosity in the hard band (2-10 keV) vs. redshift of CDF-S X-ray sources with FIREWORKS counterparts. The dashed lines mark the redshift range to which we restrict this study ($0.5<z<2.8$).
}
\end{figure}

\subsection{X-ray luminosities and nuclear obscuration}

The multi-band X-ray count rates and FIREWORKS redshifts are used
to estimate rest-frame X-ray luminosities and obscuration in the X-ray sources.
We assume that all sources have an intrinsic power-law spectrum at X-ray wavelengths with a photon index of 
$\Gamma = 1.9$ \citep{nandra94}. We simultaneously fit the fluxes in the observed-frame full (0.5-10 keV), soft (0.5-2 keV), 
hard (2-10 keV) and ultra-hard (4-10 keV) X-ray bands, corrected for Galactic absorption, to estimate 
the intervening Hydrogen column density (N$_{H}$), using the photoelectric crossections of \citet{morrison83}. 
These fits are used to derive intrinsic (de-absorbed) hard X-ray luminosities \lhard\ in the rest-frame 
2-10 keV band.

In Fig.~\ref{lx_vs_z}, we plot \lhard\ against redshift for our sample. 
Nine sources had N$_H$ estimates that were greater/lesser than the upper/lower allowed bound 
(log N$_H = 25/19$ cm$^{-2}$), typically because they are detected in only a single X-ray band.
We removed these sources from our sample. Galaxies with L$_{{2\textrm{-}10}} < 10^{42}$ \ergs\
can possibly be contaminated by X-ray emission from star formation and stellar remnants \citep{bauer02}. Therefore, we set
this luminosity limit as a minimum criterion for an X-ray source to be a pure AGN. The application of these two criteria
excludes  58 objects, leaving us with a final working sample of
118 bona fide AGN spanning a redshift range of $0.5$--$2.8$ and almost 3 orders of magnitude in X-ray
luminosity.

The de-absorbed X-ray luminosity is generally quite robust to the exact parameters of the X-ray fits, since for most
of our redshift range, the rest-frame 2-10 keV band is bracketed by the soft and hard bands in the observed frame. In
addition, the optical depth to photoelectric absorption in the hard band is low for typical obscuring
columns in X-ray selected AGNs. However, the value of N$_H$ is generally more uncertain and has systematic
variations with redshift, as the obscuration-sensitive rest-frame soft band is redshifted
out of the Chandra/ACIS band-pass. In particular, higher redshift AGNs will have systematically higher N$_H$.

\begin{figure*}[t]
\figurenum{3}
\label{mass_distribution}
\centering
\includegraphics[width=\textwidth]{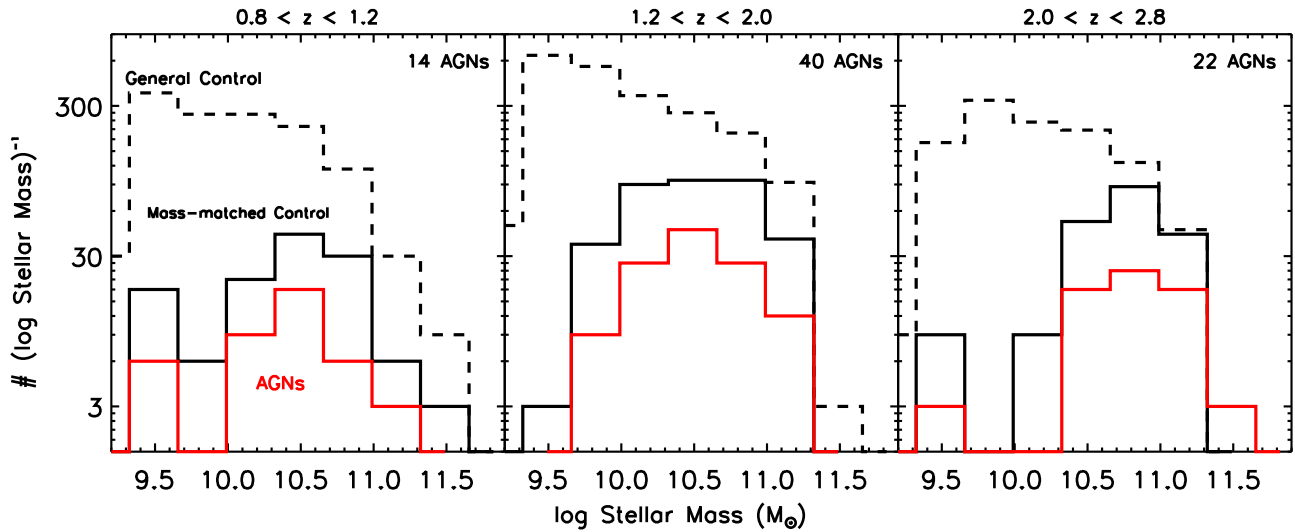}
\caption[Mass Distributions]
{ Stellar Mass (M$_*$) distributions of AGNs (red solid lines) compared to the general control sample (dashed lines)
and the mass-matched control sample (black solid lines) in three redshift bins. The AGNs (and mass-matched control
galaxies) tend to be more massive than the field galaxy population.
}
\end{figure*}

\subsection{Control sample}

The main goal of this work is to explore the nature of the host galaxies of AGN
at high redshifts. In order to fully understand AGN hosts, they must be placed within the
context of the normal field galaxy population at comparable epochs. Many studies have shown that
there has been substantial evolution in the properties of normal galaxies since $z \sim 3$: the 
star-formation rate at a given stellar mass has decreased by an order of magnitude
\citep{noeske07, tresse07},  the space density and stellar mass of red galaxies has increased
by a similar amount \citep{arnouts07, taylor09} and stellar mass functions of galaxies show 
considerable evolution, especially at the low-mass end \citep{marchesini09}.  In addition, these
studies also demonstrate considerable correlation of galaxy properties with stellar mass.
We control for these transformations and variations by defining two sets of normal galaxies, 
to which we compare the properties of AGNs throughout this paper. The first set is a representative
sample of inactive galaxies with masses and luminosities comparable to the AGNs. The second
is a refined sample with a stellar mass distribution that is matched to the AGNs.

First, we examined the images of all FIREWORKS galaxies without X-ray detections
in the redshift range $0.5<z<2.8$ which occupy the same range in absolute B-band magnitude
and stellar mass as the AGNs ( log M$_{*} > 9.3$ M$_{\odot}$ and M$_{B} < -19.5$). 
We excluded any that did not have images in the WFC3 J/H bands and 
the four GOODS ACS bands, as well as a few that lay too close to the edges of the ACS and WFC3 mosaics,
or those with image defects or contamination from bright nearby stars or galaxies. This left us with
a high quality set of 1683 inactive galaxies with a similar range of luminosity, mass and multi-wavelength coverage as the AGNs
in our sample. The redshift distribution of these galaxies, hereafter called the `general control sample',
is similar to that of the AGNs (lower panel of Fig.~\ref{kmag_vs_z}). 

From the general control sample, we applied a further selection to get a `mass-matched control sample'. First, we divided 
the AGNs and normal galaxies by redshift into the same redshift bins as those used in our subsequent comparative study (see later sections). 
For each AGN in a bin, we randomly chose three distinct normal galaxies in the same redshift bin with a stellar mass 
within 0.2 dex of the AGN's mass. This tolerance is a balance between a statistically large sample of 
control galaxies and the need to closely match the mass distribution
of the AGNs, especially at high masses, at which the AGN fraction among galaxies can be quite large at $z\sim2$.

At intermediate to high AGN luminosities, contamination of optical and IR light by the 
emission from the active nucleus can systematically alter stellar mass estimates for AGN
hosts.
In the next section, we outline a method to exclude AGNs with the strongest
contamination, using a combination of X-ray luminosity and obscuration cuts. These cuts only affects a small fraction
of objects (9\%). For most AGNs in our sample, the effects of contamination are minimal and 
do not significantly bias our control sample selections.

In Fig.~\ref{mass_distribution}, we compare the stellar mass distributions of the two control samples with the AGNs
in three redshift bins: $0.8$--$1.2$, $1.2$--$2.0$ and $2.0$--$2.8$. These three bins will be used frequently
in the rest of this work and, for brevity, are termed the `low', `intermediate' and `high'  bins respectively.

The general control sample has a mass distribution (dashed line) that is approximately Schecter-like, while that of the
AGNs and mass-matched control sample (red and black solid lines) drop-off strongly towards low masses. 
Note the close correspondence between these latter two distributions, which highlights the quality of our mass-matching
procedure.

AGNs are weighted towards being found in rather massive galaxies. The peak in the 
AGN mass distribution is at $\log M_{*} \approx 10.5-10.7$, increasing slightly with redshift. 
In addition, the fraction of massive galaxies hosting AGNs increases drastically with redshift. While AGNs are only
a few percent of similar mass galaxies in the low bin, the fraction increases to more than 30\% at the high mass end
in the high bin. 
The typical stellar masses of the AGN hosts are much higher than the mass incompleteness limit of the FIREWORKS catalog at all
redshifts. 
This implies that the slight evolution in the peak
mass is probably real and not a consequence of differences in the AGN luminosity range with redshift. It may be related
to the strong increase in the space density of AGNs. A detailed look at mass functions of AGNs is beyond the scope of this 
paper, but we discuss some aspects of the evolution of the AGN mass function in \S10.

\section{Aperture Photometry}

\begin{figure}[ht]
\figurenum{4}
\label{psf_scatt_test}
\centering
\includegraphics[width=3.5in]{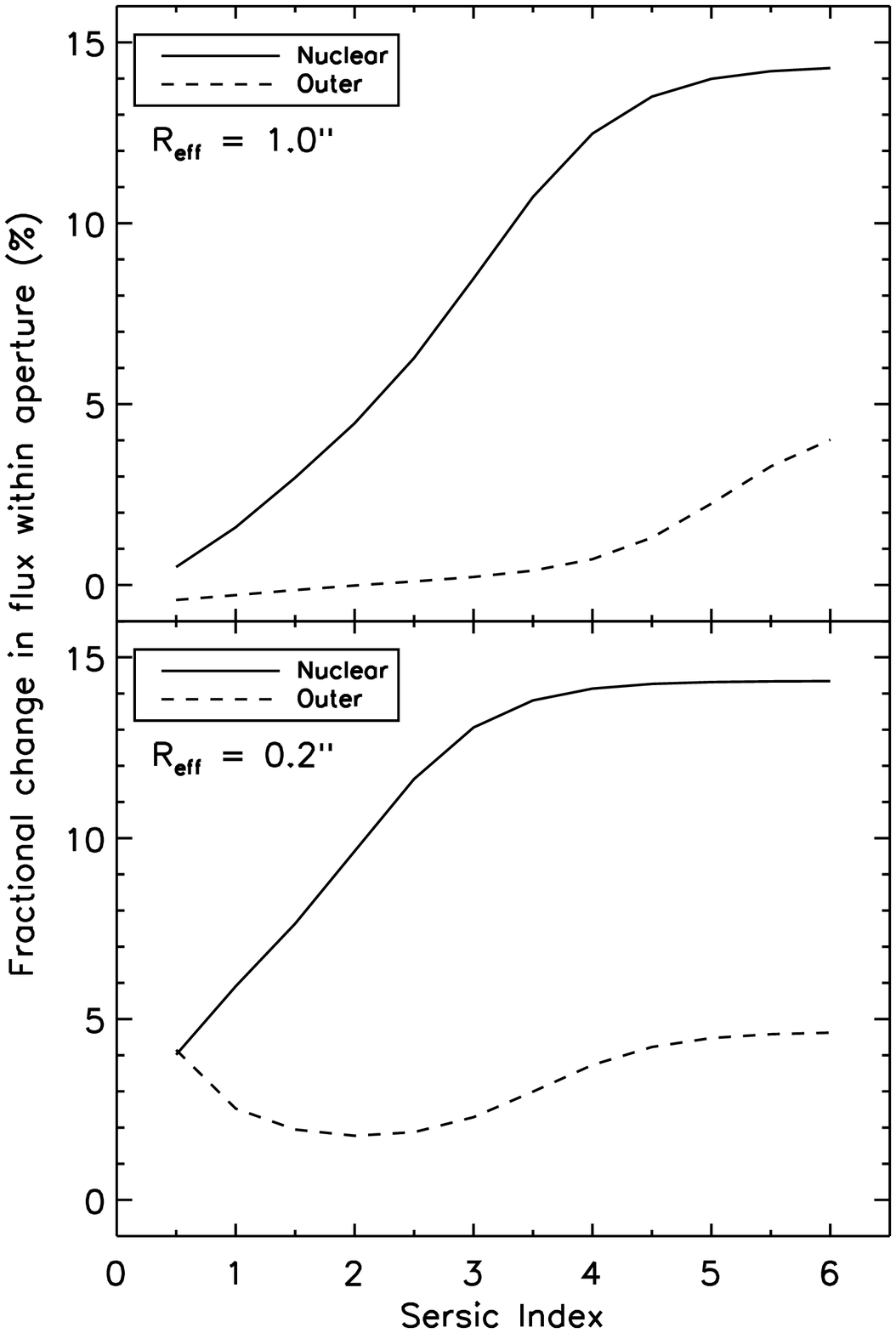}
\caption[PSF scattering test]
{ The absolute fractional difference between the flux in simulated galaxies convolved by two different models of the
CANDELS/H-band PSF, as a function of the original S\'ersic index of the simulated galaxy. The two model
PSFs are the drizzled TinyTim CANDELS-Deep PSF, an accurate model of the actual PSF, and a GOODS/ACS
B-band PSF, matched to the H-band using PSFMATCH kernels. The fluxes are measured in two apertures: a 
nuclear aperture with a radius of $0\farcs1$ and an annular `outer' aperture of radius $0\farcs4$--$1$".
The difference between the fluxes are expressed as a fraction of the flux in the aperture from the CANDELS-Deep
PSF convolved image. The panels show results for two sets of simulated galaxies with different \reff: very
compact galaxies in the lower panel and relatively large galaxies in the upper panel. The nuclear aperture is affected
considerably by light scattered from the core by imperfections in our PSF-matching procedure, but the outer aperture
is only marginally affected by scattered light ($<5$\%). 
}
\end{figure}

For each galaxy in our AGN and control samples, we performed aperture photometry on
the full set of seven HST images from 4 ACS/GOODS and 2-3 WFC3/CANDELS bands.
We first used the PSFMATCH routine within IRAF \footnote{Image Reduction and Analysis Facility, distributed by the National Optical Astronomy
Observatory} to develop convolution kernels to match the PSFs of the images in all bands to those of the H-band. 
The PSF-matching procedure reproduces the extended structure of the H-band PSF very accurately in all the GOODS and WFC3 images.
However, the simulated PSFs had slightly broader cores than that
in the H-band (by 15-20\%, depending on the width of the original PSF), a result of the apodizing filter applied
to the convolution kernels to remove high-frequency noise in the original PSFs.  
Photometry of galaxies in apertures close in size to that of the WFC3 H-band PSF (such as the
`nuclear' aperture used in later sections) yield SEDs that are too red, as some of the light in the core of the
bluer bands is scattered out of the aperture by the PSF-matching process. The degree to which this effect operates
is a strong function of the compactness of the galaxy image. Fig.~\ref{psf_scatt_test} illustrates
how differences in the shape of simulated PSFs affect the photometry of galaxies.
We convolved a suite of circular S\'ersic model galaxies (described in the Appendix) with two PSFs: the drizzled TinyTim
H-band PSF  and the ACS B-band PSF matched to that of the H-band using our PSFMATCH kernels. 
The B-band was chosen for this illustration since its PSF differs the most from the H-band PSF. 
In the Figure, we plot the fractional difference between the flux of a model galaxy convolved by the two PSFs as a 
function of the S\'ersic index of the model. The two panels represent two
sets of simulated galaxies with very small and relatively large \reff. 

In general, imperfect PSF matching leads to larger differences in the nuclear aperture (as much as 14\%)
but rather negligible effects in the default outer aperture ($< 5$\%, but only for high S\'ersic galaxies).
This holds for both the most compact as well as fairly large galaxies.
Clearly, the artificial `PSF-reddening' will not affect the results of our extended light photometry (\S7),
but may play a role in the interpretation of color gradients (\S8), as discussed in that section.


After convolving the images in each band to bring them to a common PSF, we used the IDL program APER 
to measure counts in a set of annular apertures, ranging from $0\farcs1$ to $4"$, centered on the coordinates of the the 
H-band centroid. The photometric measurements were converted to AB magnitudes using zero-points taken from
GOODS v2.0 \citep{giavalisco04} and STScI WFC3 documentation. 
Errors were estimated from the multidrizzle weight maps, with a standard correction for correlated pixel-to-pixel noise. 

\begin{figure*}[ht]
\figurenum{5}
\label{nuclear_colors}
\centering
\includegraphics[width=\textwidth]{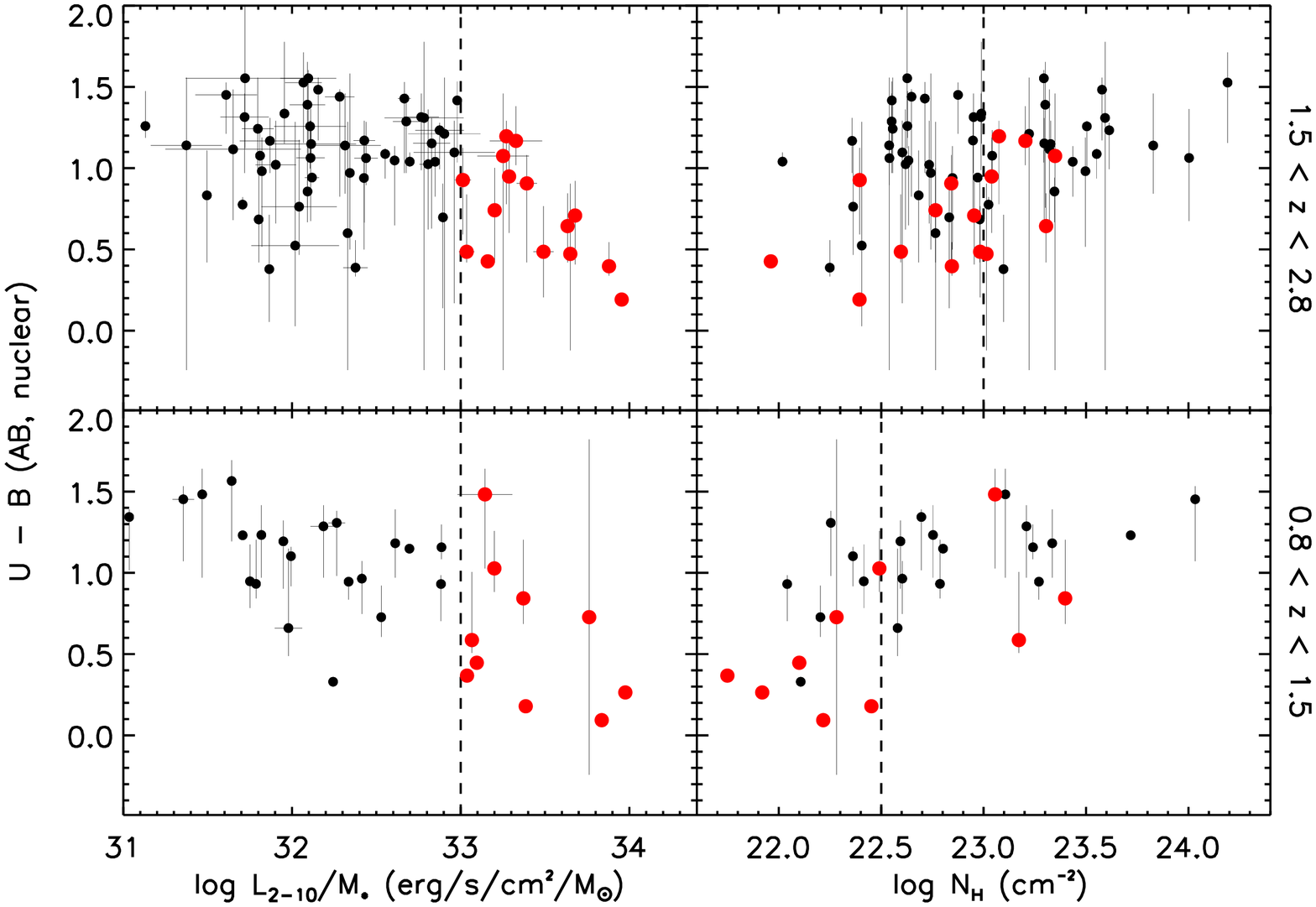}
\caption[Nuclear colors against X-ray parameters]
{ The rest-frame nuclear U-B colors of AGN hosts plotted against the ratio of 
hard-band X-ray luminosity to stellar mass (left panels) and the Hydrogen column density of X-ray obscuring
gas N$_H$ (right panels). The lower panels show AGNs at intermediate redshifts ($0.8<z<1.5$), while the upper panels
show AGNs at high redshifts ($1.5<z<2.8$). The
vertical dotted line in the left panels mark the transition value of \lhard/\mass\
beyond which nuclear emission from the AGN prominently affects the nuclear colors. 
These sources are marked as red points in all panels and appear to separate out in the plane
of color vs. N$_H$. The blue sources are generally soft, while
the red sources are typically harder, which high values of N$_H$. The dashed vertical lines
in the right panels mark the values of N$_H$ used in combination with the transition \lhard/\mass\
to separate out sources with nuclear contamination.
}
\end{figure*}

\section{Model Fits and Statistics}

\subsection{SED Fitting and Parameter Estimation}

The main thrust of this paper is to self-consistently study rest-frame properties (colors, mean extinctions) 
and the stellar populations (ages, SFHs) of X-ray selected AGN hosts across redshift, and compare them to the
properties of inactive galaxies.  Towards these ends, we employ a library of synthetic SEDs derived 
from the population synthesis models of \citet{maraston05}. Starting with a set of Single Stellar Populations (SSPs)
with metallicities of 1/2, 1 and 2 times solar, we generated models with three possible star-formation histories (SFHs) --
constant star-formation, exponentially decaying star-formation (negative `$\tau$-models') and exponentially
\emph{increasing} star-formation (positive `$\tau$-models'). The latter set of models are rarely employed in studies to
date, but recent work suggests that SFHs that increase with time may be more relevant at $z\gtrsim2$, since the mean
star-formation rate density of the Universe turns over at that epoch \citep{maraston10}. 
The star-formation history is parameterized 
by a finely sampled set of 43 ages (since the onset of star-formation), ranging from 0.01 to 15 Gyrs, 
and 16 exponential time-scales ($\tau$), ranging from 0.05 to 20 Gyr. 
In addition, we applied a foreground-screen dust reddening prescription with a 
Calzetti reddening law \citep{calzetti94}, parametrized by the visual extinction 
$A_V$ in 16 uniformly logarithmic steps from 0.001 to 7. We derived model magnitudes for each 
SED at a set of redshifts between 0.4 and 4.0, with uniform steps in redshift of 0.01. At any redshift, only models that were younger
than the age of the Universe were considered. Magnitudes were evaluated in the observed frame for the 7 HST bands,
ISAAC K$_S$ and the four IRAC bands, using the most up-to-date filter functions publicly available. Our library of model SEDs is sufficient to 
capture the bulk of spectral variation among galaxies, verified by comparing the empirical range of rest-frame colors 
among  galaxies in the FIREWORKS catalog with that of our model set.

There is some debate in the literature about the relative importance of thermally-pulsating AGB (TP-AGB)
stars on the determination of stellar populations in galaxies \citep{maraston06, kriek10, pforr12}.
Our choice of SSP models allow a strong contribution from this transient population, while other
popular models, such as those of \citet{bc03}, do not. Analysis of real galaxy samples 
show that the choice of model library can contribute to much of the systematic differences found among
studies \citep{kannappan07, muzzin09}, as well as different choices of SSP parameters such as the choice
of mean metallicity, extinction law, and the shape and turnover of the initial mass function \citep{conroy09, pforr12}.
Since we use the same modelling methodology to fit all our galaxies, comparisons between different samples 
should be internally consistent, though a comparison to other studies with a different choice of synthesis 
models will likely show systematic differences in derived SFH parameters at the factor of $2-3$ level \citep{muzzin09}.
U-V colors are unaffected by the choice of model library.

For each galaxy and a given annular aperture, we fit for a scaling parameter that minimized 
the $\chi^{2}$ difference between the model and measured fluxes, at a grid redshift closest to the measured redshift 
of the galaxy. This minimum $\chi^{2}$ for the model was then stored. 
In this way, we obtained a four-dimensional $\chi^{2}$ space for each galaxy, parametrized by SFH model (constant/range of
positive and negative $\tau$), age, $A_V$ and metallicity. We applied
an additional constraint that the synthesized K$_S$ and IRAC magnitudes for 
any model fit to the SED of an annular aperture must be less than the total measured 
K$_S$ and IRAC magnitudes of the galaxy, up to a rest-frame wavelength of 1.6$\mu$. 
In practice, this constraint only excludes the most
dusty models at the highest redshifts, but is included for physical consistency.

From this $\chi^{2}$ space, we developed a method to derive the best-fit value and accurate confidence
intervals for any model-dependent parameter (age, $tau$, rest-frame colors, $A_V$), similar to the procedure
described in \citet{salim05} for population fits to galaxies SED from UV-to-IR SEDs. 
Consider, as an example, the rest-frame U-B color of the galaxy. We concentrate on this color
since it is  uniquely sensitive to the mean light-weighted stellar population of a galaxy, 
since the two bands constituting the color lie of either side of the $4000$ \AA\ break. 
Each model has a particular value of rest-frame U-B, calculated
directly from its synthetic spectrum. The best-fit value of U-B corresponds to the model with the
lowest $\chi^{2}$ value. The uncertainty in U-B is given by the minimum and maximum color of all
models that deviate from the lowest $\chi^{2}$ by less than 3.5, which defines the appropriate
68\% ($1\sigma$) confidence interval for three degrees of freedom (7 bands - 4 model parameters). 
The advantage of using the full  $\chi^{2}$ space is that systematic uncertainties due to 
degeneracies in the model space, which can frequently be larger than purely statistical uncertainties, 
are taken into account in this method.

In this way, we arrive at best-fit values and confidence intervals for observed and intrinsic rest-frame colors,
visual extinction, stellar age, $\tau$ and the ratio of stellar age over $\tau$ (the `normalized age') for each
galaxy. The last two quantities are only appropriate for galaxies fit with $\tau$ models. Note that systematic
differences may be expected in the best-fit values of SFH 
parameters depending on the choice of synthesis library. A detailed treatment of these systematics is beyond the
scope of this work. However, since we use the same library for all our fits, comparisons between
AGNs and control galaxies should be internally consistent.

\begin{figure*}[ht]
\figurenum{6}
\label{qso_stamps}
\centering
\includegraphics[width=\textwidth]{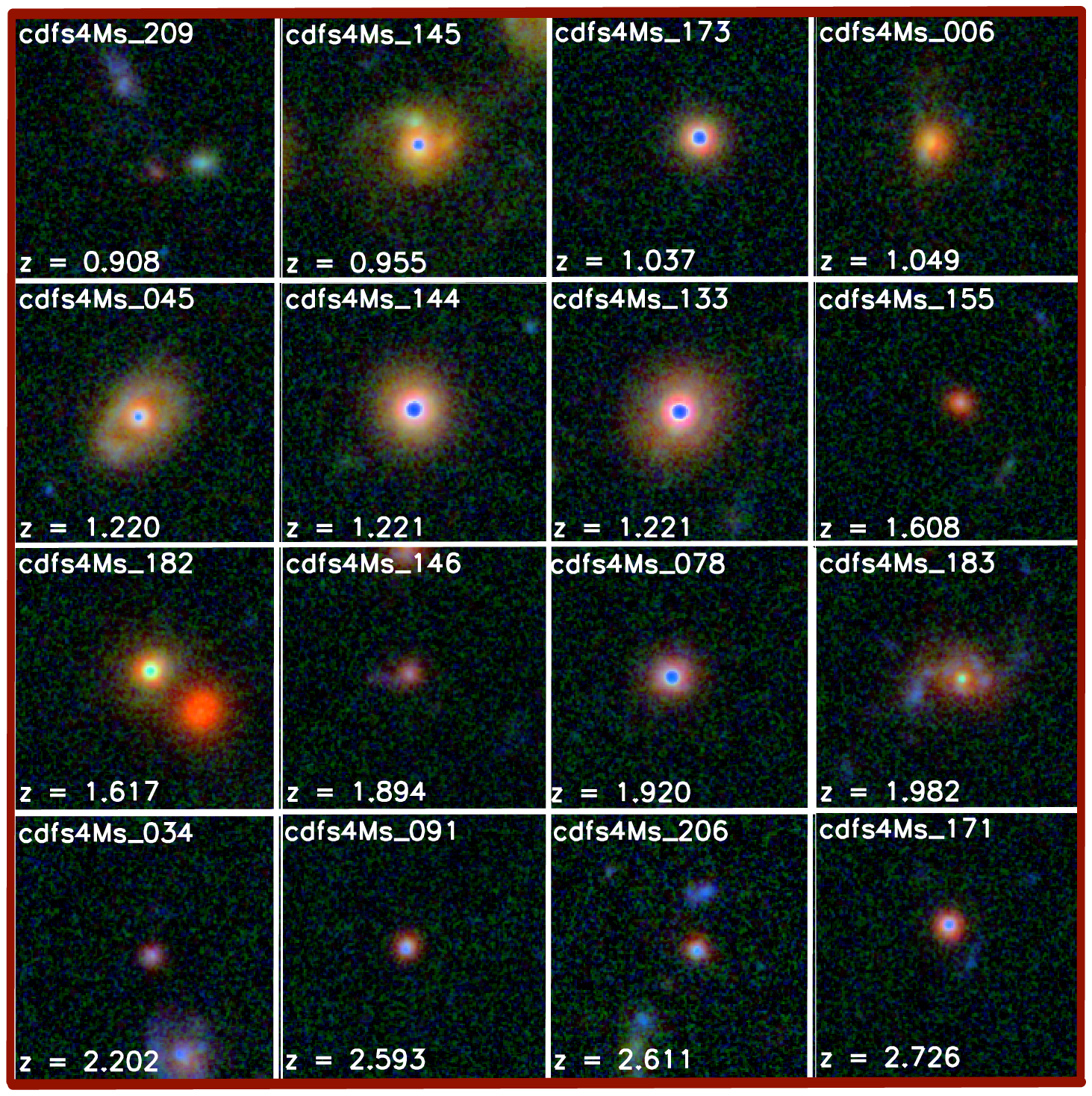}
\caption[Objects with Nuclear contamination]
{ Three-color VzH images of the sixteen AGNs in the redshift range $0.8<z<2.8$ which have been flagged 
as galaxies with strong nuclear AGN contamination, based on X-ray based criteria (\S6). 
Bright blue nuclear point sources are seen in almost all these sources.
}
\end{figure*}

\subsection{Monte-Carlo Bootstrapped Distributions}

A important part of the following analysis involves comparing distributions of various estimated quantities (colors,
SFH parameters) between AGNs and inactive control galaxies. The small size of the sample and the large
uncertainties on some of the parameters from the SED fitting analysis makes interpretation of any differences
in the distributions of best-fit parameters rather difficult. As 
the library of models do not uniformly sample parameter space, best-fit values of a parameter can be
over-represented in regions of parameter space that are more well-sampled than others. 

To enable a reasonable comparison to be made between the parameter distributions of
AGNs and control galaxies, we developed a two step monte-carlo bootstrap approach to arrive at 
more representative distributions for both samples. For each object, we take the best-fit parameter from the SED fits
to be a median value. We randomly vary the parameter about this value
taking the upper and lower $1\sigma$ confidence intervals as the standard deviations of a two-sided 
piece-wise gaussian probability distribution, where the probability of a random deviate being positive or negative 
is 0.5 (by construction, as the best-fit value is assumed to be a median value). For some parameters, such as stellar age,
we apply upper and lower bounds to the random deviate consistent with any limits enforced in the SED fitting process. 
This process is repeated 1000 times per object. The final distribution in each redshift bin is derived from
the entire simulated set of parameters for all objects (AGNs or control) in that bin. We term this bootstrapping step `Loop 1'. While 
Loop 1 takes into account the uncertainties that come out of the SED fitting process, it does not
consider the effects of the limited sample size of the AGNs -- stochastic effects can play a big role in the shape of
a parameter distribution. To account for this, we run Loop 1 on a subsample of $N$ randomly chosen galaxies from the control 
sample in a redshift bin, where $N$ is the number of AGNs in the same redshift bin. This second set of monte-carlo 
simulations is called `Loop 2' and it allows us to place upper and lower $1\sigma$ uncertainties on the distribution of the control galaxies 
resulting from stochastic effects from the small size of the AGN sample in any given bin.
The distributions of AGNs and control galaxies are only significantly different if they have consistent offsets
that are larger than the $1\sigma$ errors bars on the control galaxy distributions. Note, however, that this
process does not regularize stochastic variations due to the limited sample size of the control galaxies.
In all redshift bins, the mass-matched control galaxy samples range between 45--110 objects, and the general control sample,
of course, is much larger. Therefore, stochastic effects among these galaxies
should be small. However, we have checked the validity of our main results through several iterations of the
mass-matching procedure described in \S3.3. 
All our conclusions are essentially unchanged in several randomly different mass-matched control samples.

\section{Nuclear Colors and Obscuration}

The active nucleus is a source of considerable high-energy radiation. In cases where the emission from the AGN
is relatively unobscured to optical light, the nucleus can appear as a point-source in galaxy images, 
frequently blue in color. Bright unobscured nuclear point sources can influence the integrated colors of AGNs, and, 
quite drastically, the apparent morphology of AGN hosts \citep{pierce10a}.
The high spatial resolution of the WFC3 NIR images allows a better characterization of the 
SED of the nuclear light in AGNs at $z\sim 1$ and allows us to constrain
the contribution of non-stellar emission to the nuclear light in AGNs at $z\sim 2$ more accurately than ever before.

The relative degree to which light from an unobscured active nucleus can contaminate the light of a galaxy is a related directly  
to the luminosity of the AGN and inversely to the stellar luminosity of the galaxy. To account for both dependencies,
we construct a parameter \lhard/\mass\ which is the ratio of the X-ray luminosity of an AGN to this stellar mass of its host.

In Fig.~\ref{nuclear_colors}, we plot the rest-frame `nuclear' U-B color of the X-ray AGNs against \lhard/\mass\ 
(left panels) and the X-ray obscuring hydrogen column density N$_H$ (right panels).
The `nuclear' photometry is estimated within a central circular aperture $0\farcs2$ in diameter, corresponding to a physical
scale of $\approx 1$ kpc across our redshift range of interest. The aperture size is larger than the $0\farcs15$
FWHM of the WFC3 H-band PSF and encloses $\approx 40$\%\ of any unresolved central light. 
We separate the AGNs into two redshift bins in this plot: $0.8<z<1.5$ and $1.5<z<2.8$.

The nuclear colors of AGNs at $0.8<z<1.5$ are predominantly redder than $\textrm{U-B}=0.8$ (approximately the
top of the Blue Cloud at these redshifts -- see \S7.3.1). However, above a transition value of $\log \textrm{\lhard/\mass} = 33$
\ergs M$_{\odot}^{-1}$ (red points), some AGNs show scatter towards much bluer nuclear colors.
In addition, the blue color appears to correlate with \lhard/\mass, suggesting that the excess blue light in the centers
of these AGNs comes from nuclear emission. Indeed, a visual inspection of these bright AGNs with blue nuclei
clearly show nuclear point sources (Fig.~\ref{qso_stamps}). Most also have broad AGN lines visible in their spectra \citep{szokoly04}.

The right panels of Fig.~\ref{nuclear_colors} allow us to compare the X-ray obscuration properties of the AGNs as a function of nuclear
color. In the lower right panel, the red and blue branches of the bright AGNs (red points)
separate out in N$_H$. Objects with blue colors all show relatively low obscuration (log N$_H \sim 22$), while those
with red colors show a larger range in N$_H$, but tend to be higher. In other words, luminous
AGNs that are unobscured towards the nucleus in optical light are also relatively unobscured at X-ray wavelengths,
while those that are obscured at optical wavelengths generally show a larger X-ray obscuration as well.  

The anti-correlation between visible nuclear optical emission and the level of X-ray obscuration has been noted in 
previous work on AGN host color gradients, typically at $z\sim 1$ \citep{pierce10b}. The simplest interpretation of 
this trend is the existence of a relationship between the medium that obscures the optical light of the AGN 
accretion disk and the medium that obscures the X-ray emission. 

A similar behavior is seen at high redshifts (top left and right panels), with some qualitative differences. The 
colors of the unobscured AGNs are not as blue as at lower redshifts. This could be because the optical
obscuration towards the nucleus at $z\sim2$ is larger on average compared to that at $z\sim1$, since the gas fraction of
galaxies increases towards high redshifts. Another possibility is that the fraction of stellar light within the nuclear aperture is
larger at $z\sim2$, as galaxies at these redshifts are generally quite compact. This will reduce the contrast of AGN 
emission against that of the galaxy.

In addition, the mean N$_H$ between the AGNs with blue and red nuclear colors
is not as pronounced at $z\sim2$ as at intermediate redshifts. This difference is hard to interpret 
accurately, since the X-ray spectral fit parameters, and especially N$_H$, can vary systematically with redshift. 
If true, it may represent a gradual change in the relationship between optical and X-ray obscuration towards
higher redshifts. For example, at $z\sim1$ optical obscuration could come primarily from dusty gas in the vicinity of the
SMBH, such as the putative dusty torus which is invoked in Unification schemes for local AGNs \citep{antonucci93}.
However, at $z\sim2$, more of the optical obscuration may arise from intervening dust within the gas-rich ISM of
the host galaxy. In both cases, most of the X-ray obscuration probably arises in very dense gas surrounding
the AGNs accretion disk, since the typical gas mass surface densities of galaxy disks generally 
cannot account for the high values of N$_H$ \citep[e.g.,][]{kennicutt12}. 

We are now able to decide on criteria to allow us to flag and remove AGNs with likely 
strong nuclear point sources from the following analysis of outer
colors and star-formation histories. Since nuclear colors can be blue due to both AGN emission or 
strong star-formation, we choose not to use pure color criteria, since this can remove 
galaxies with genuine nuclear star-formation, rather than strong AGN activity. 
Instead, we employ purely X-ray based criteria. We decide on log  $\log \textrm{\lhard/\mass} = 33$
\ergs M$_{\odot}^{-1}$ and log N$_H < 22.5/22.9$ cm$^{-2}$
at z lesser/greater than 1.5, guided by the results of Fig~\ref{nuclear_colors}. 

\section{Host galaxy colors}

In this section, we study the rest-frame U-B color of the AGN hosts in annular apertures that are relatively
free of nuclear contamination, and compare them to the rest-frame U-B color of various control galaxy samples
across various redshift bins. 


\subsection{Choice of Working Annulus}

In this study of extended light, we will work in an annular aperture of fixed angular size across all redshifts. This is preferred over working with apertures of fixed physical size for two primary reasons:

\noindent a.) Any comparison of photometric properties between annular apertures of different sizes will need to
take into account the smearing effects of the PSF, which is dependent on the light profile of the galaxy, strength of nuclear
AGN emission and the radius of the annular aperture. 
These effects can be hard to model {\it a priori} and may have considerably uncertainties
and systematics. However, working in fixed annular apertures at all bands and redshifts keeps these effects
fixed as well, and greatly simplifies the interpretation of any comparisons.

\noindent b.) The angular diameter distance scale varies by less than 15\%\ between the redshifts of 0.8 
and 3 and only by about 30\%\  between redshifts of 0.5 and 1. Therefore, a comparison of outer colors 
of galaxies between $z\sim1$ and $z\sim2$ at fixed angular apertures is essentially equivalent to a 
comparison at fixed physical apertures. 

To arrive at an optimal annular aperture to study the extended light of AGN hosts, 
we rely on the measurements of the sizes of AGN hosts (and inactive galaxies) at $z\sim2$
presented in \citep{schawinski11}. This work used the two-dimensional profile modeling code
GALFIT \citep{peng10} to model the H-band images of high redshift
AGNs (including the effects of central point sources) in the ERS2
field and find that their half-light radii \reff\ span a typical range from $0\farcs25$-$0\farcs75$, which
corresponds to approximate physical sizes of 2-6 kpc. 

For our purposes, we choose an outer annular radius of 1", which encompasses a substantial fraction of the 
light of galaxies across our redshift range of interest. The inner radius of the annulus 
has to be large enough to minimize contamination from the central light of the galaxy (nuclear or stellar), 
so that valid measure of color gradients can be made. After experimenting with the relative size of the 
H-band PSF and various aperture sizes using a suite of model AGNs (see Appendix), we settled on an inner aperture of $0\farcs4$. 
Beyond this radius, the extended blue light in the smaller galaxies is contaminated by unresolved central emission
at the level of a few tens of percent, and even less for the larger galaxies. 

Having said this, nuclear continuum emission in bright AGN can still substantially affect 
the extended light aperture at blue and UV wavelengths, which will skew our SED fits towards bluer colors 
and younger mean ages. Therefore, we also employ the criteria based on our study 
of nuclear colors to remove the small fraction of AGNs with strong nuclear point sources.

\begin{figure*}[t]
\figurenum{7}
\label{extended_cmds}
\centering
\includegraphics[width=\textwidth]{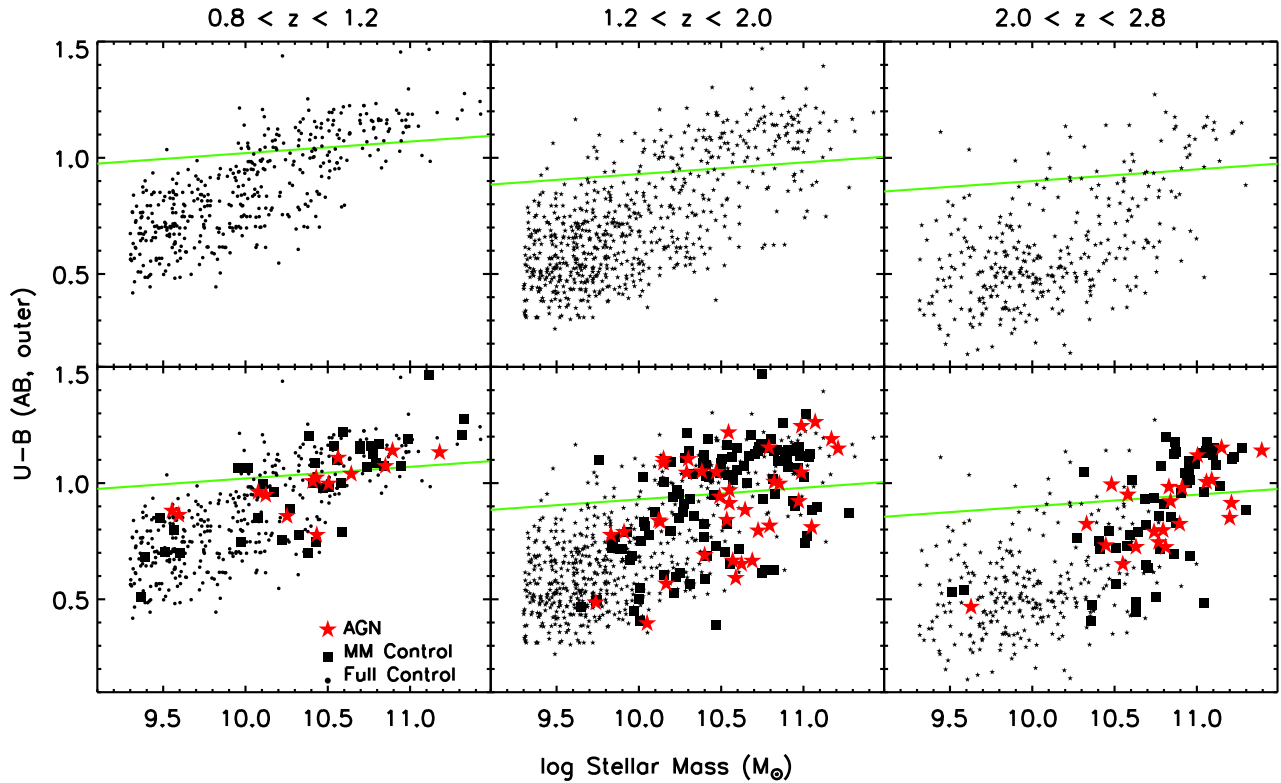}
\caption[Extended CMDs]
{  Outer U-B color vs. stellar mass for AGNs and normal galaxies in three redshift bins. In the upper
panels, the small points show all galaxies in the FIREWORKS catalog with  log M$_{*} > 9.3$ M$_{\odot}$
and M$_{B} < -19.5$ (the general control sample), for which we have performed extended light photometry. 
The green lines mark
the location of the Green Valley, derived separately in each bin. In the lower panels, in addition to the
same small points as in the upper panels, the large points show outer colors
for AGNs (red) and mass-matched control galaxies (black). The AGNs tend to have high 
stellar masses and redder colors than the normal galaxy population.
}
\end{figure*}

\subsection{Context: Field Galaxy Color-Mass Diagrams}

Before a detailed study of AGN colors, we set the context by examining the extended light colors
of inactive galaxies through the use of U-B color-mass diagrams (\cmd).
In the upper three panels of Fig.~\ref{extended_cmds}, we plot the \cmd\
of galaxies in the three redshift intervals defined earlier: low ($0.8<z<1.2$), intermediate ($1.2<z<2.0$) and
high ($2.0<z<2.8$) bins. U-B colors in this Figure come from extended light photometry,
which generally differ integrated colors.

We begin by highlighting the bimodality evident in the outer colors of the non-active control galaxies at all redshifts. 
A Red Sequence from evolved or reddened galaxies can be easily distinguished in the low and intermediate bin panels.
A slope to the sequence is discernible, though the data are not sufficient to identify any evolution in the slope.
In the high bin, the low-mass end (log M$_{*} < 10.5$) of the Red Sequence is diminished, but red galaxies can still
be found at high masses. 

In addition to the Red Sequence, a cloud of blue, star-forming galaxies is also
seen in the \cmd, typically at lower masses than the red galaxies. The density of the Blue Cloud
relative to the Red Sequence, increases with redshift, while the typical color of the Blue Cloud decreases with redshift:  
median U-B values for low/intermediate/high bins are 0.7/0.6/0.45 mag. This is primarily 
due to higher star-formation and lower characteristic stellar ages of star-forming galaxies at higher redshifts.
Since we deal primarily with AGN host properties in this paper, we refrain from a detailed treatment of the 
form of the bimodality and its evolution with redshift, noting only that a bimodality exists and is well-defined. 


\begin{figure}[t]
\figurenum{8}
\label{colors_masses}
\centering
\includegraphics[width=3.5in]{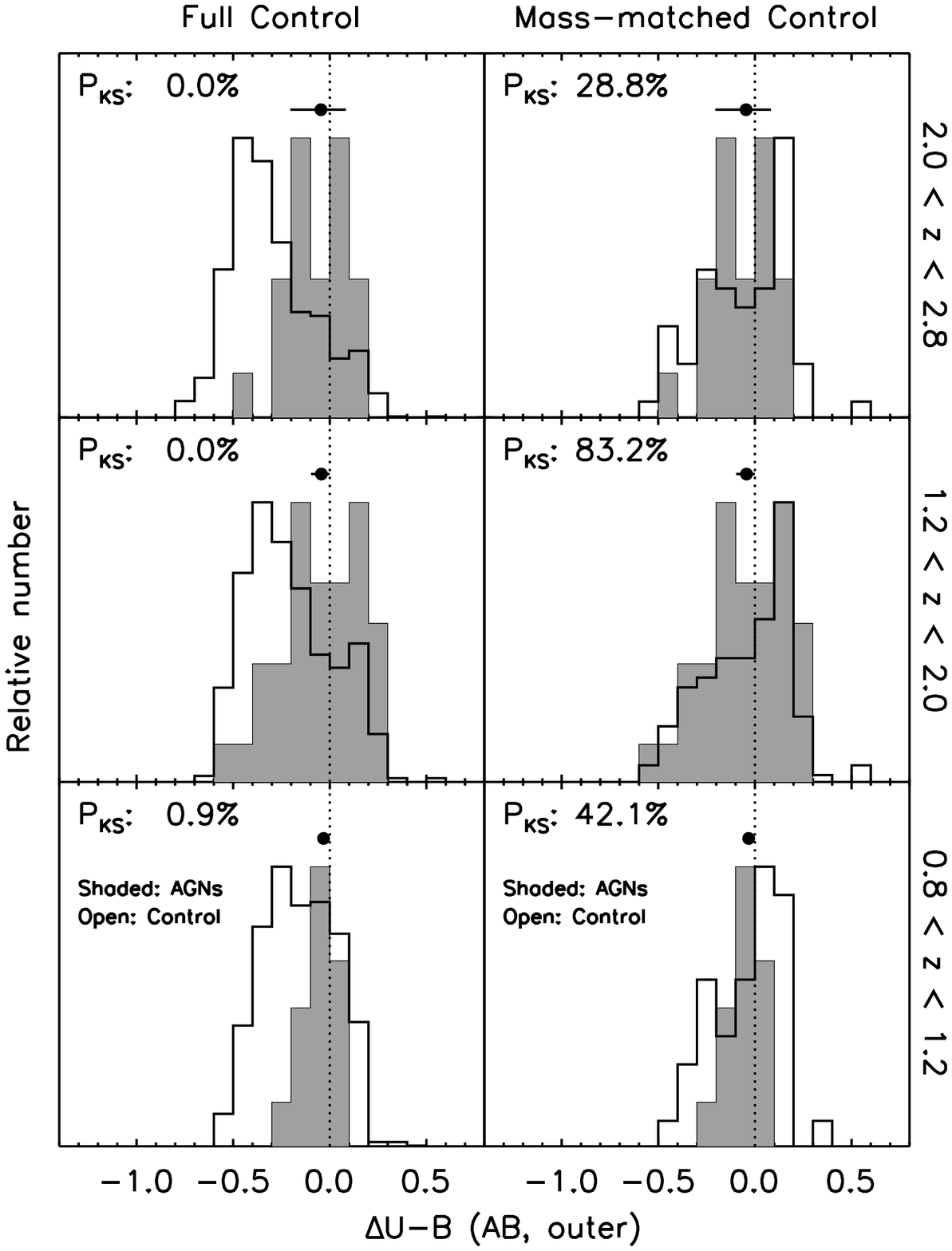}
\caption[Extended Colors and Mass distributions]
{  Distributions of Outer U-B color offsets from the Green Valley measured in annular apertures of $0\farcs4-1\farcs0$
in three redshift bins, increasing in median redshift from bottom to top. The left panels show the U-B distribution
with respect to the general control sample, while the right panels are a comparison to the mass-matched
control sample. The Kolmogorov-Smirnov probability P$_{\rm{KS}}$ is a measure of the likelihood that the
two distributions are derived from the same parent distribution. The dashed line marks $\Delta$U-B$=0$, i.e,
the location of the Green Valley. Error bars above the histograms give the median error in the $\Delta$U-B
of the AGNs.
}
\end{figure}
%

We use the \cmd\ to evaluate the location of the Green Valley in extended light. 
Since radial color gradients exist among massive star-forming galaxies \citep[e.g.][]{guo11}, 
we expect the location of the Green Valley in these \cmd\ to differ slightly from definitions based on integrated 
photometry \citep[e.g.][]{willmer06}. We therefore define our own Green Valley
as the dividing line between the Red Sequence and Blue Cloud, corresponding to the region of minimum
density in the \cmd\ between these two populations of galaxies. Since the Red Sequence has a definite
slope with stellar mass, the Green Valley line will also be sloped. 
We assume that the 
slope of the Green Valley in our U-B \cmd\ 
is constant with redshift. We calibrate this slope in the intermediate bin, which has the largest density of galaxies, 
by varying the line parameters to place it at a minimum in the distribution of 
U-B offsets from the line. We derive a slope to the Green Valley line of 
0.09 mag per dex in M$_{*}$. For the other two redshift bins, we kept the slope constant at this value, 
but vary the intercept to identify the Green Valley.
The location of the Green Valley is shown in each panel of Fig.\ref{extended_cmds} with solid green lines.

\subsection{AGN Hosts vs. Inactive Field Galaxies}

The upper panels set the context of the field galaxy population within which we couch the 
\cmd\ of the AGN hosts. In the lower three panels of Fig.\ref{extended_cmds}, 
we plot the AGNs as red stars and mass-matched control 
galaxies as black squares, in addition to the general control sample.  

At all redshifts, the AGNs are distributed differently in the \cmd\ as compared to the general control sample. 
Their stellar masses are typically higher and their colors are typically redder, and they display
a flatter distribution of U-B colors with a weaker bimodality.
This tendency has been observed in many previous studies of the integrated photometry of AGNs \citep[e.g.,][]{nandra07, cardamone10},
and we show here, using extended light photometry, 
that it is not due to low levels of nuclear blue light pushing red hosts into the 
Green Valley, but is intrinsic to the stellar populations of the AGN hosts.

A clearer picture of the differences between AGNs and field galaxies sample comes from their
distribution of U-B offsets.  The offset is defined as the difference in the
U-B color of a galaxy from the Green Valley color at its particular stellar mass, i.e, it is the offset of the galaxy from the
Green Valley line defined above. We refer to this offset as \dcol. 

In the left three panels of Fig.\ref{colors_masses}, we compare the \dcol\ distributions of AGNs (solid histograms)
and the general control sample (solid histograms) in all three redshift bins. These histograms show the best-fit colors determined from the SEDs and have not been `regularized' by the bootstrapping procedure described in \S5.2. 
Nevertheless, the difference between the distributions is clear. 
The general control sample shows a wide range in \dcol, with a peak at a negative (blue) value that decreases
with redshift. The AGNs, on the other hand, display a narrower distribution that peaks at significantly
redder colors, more or less around the Green Valley (\dcol$=0$). The distribution of AGN colors is narrowest
in the low bin and gets progressively broader in the two higher redshift bins. Note that this broadening
is unlikely to be due to a larger errors in the photometry of higher redshift galaxies. The median errors in
the \dcol\ of the AGNs are shown above the histograms and are generally much smaller than the widths
of the color offset distibutions. 

We compare the distributions of the control sample and AGNs 
using a Kolmogorov-Smirnov (KS) test, which gives us the probability P$_{\rm{KS}}$ 
that the two distributions are drawn from a common parent distribution. P$_{\rm{KS}}$ for AGNs compared to 
the general control sample are all less than 1\% in all three redshift bins.
 
\begin{figure}[t]
\figurenum{9}
\label{color_trends}
\centering
\includegraphics[width=3.5in]{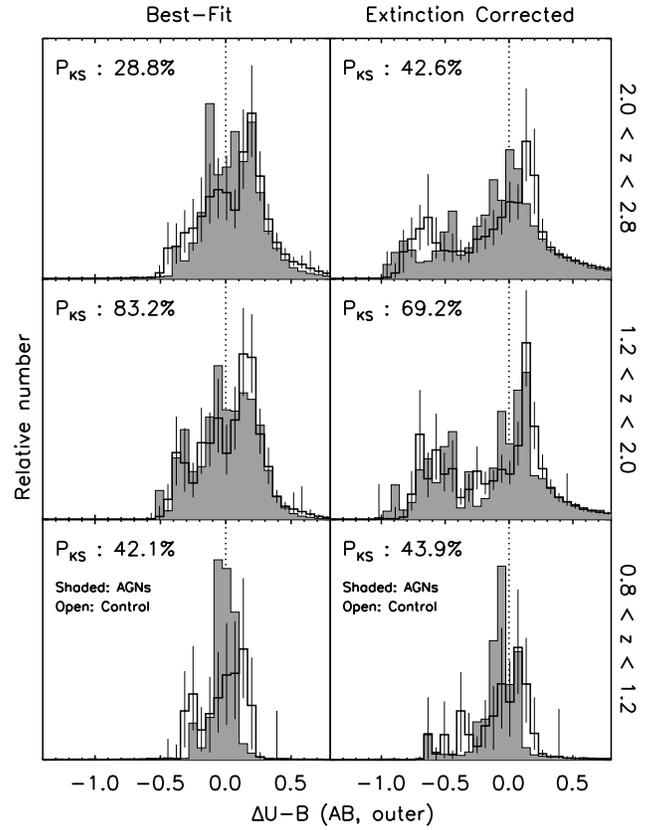}
\caption[Trends in Outer colors]
{  Distributions of Outer U-B color offsets from the Green Valley measured in annular apertures of $0\farcs4-1\farcs0$
in three redshift bins, increasing in median redshift from bottom to top. AGNs are represented by a shaded
histogram, while the mass-matched control sample is represented by an open histogram. Error bars on the 
distributions are combinations of model uncertainties and small sample statistics of the AGNs.
The left three panels on the left show the distributions
of best-fit color, while the panels on the right show the distributions of color for best-fit SED models before applying any
extinction (in essence, an extinction-corrected color of the galaxy). These distributions have been
regularized by a monte-carlo bootstrap procedure (\S5.2). The dashed line marks $\Delta$U-B$=0$, i.e,
the location of the Green Valley. 

}
\end{figure}

\subsection{Stellar Mass Selection Effects}


We are in a position now to tease apart stellar mass selection 
effects from factors related to SMBH accretion among X-ray selected
AGNs. The right set of panels in Fig.\ref{colors_masses} compares \dcol\ distributions of the AGNs (filled histograms)
and the mass-matched control sample (open histograms) in the three redshift bins. A simple comparison of the middle
and right panels shows that the mass-matched control sample is a much better match to the color distribution of the AGNs
than the general control sample. In particular, the range in \dcol\ and the peak value of the mass-matched histograms
are now quite similar to the AGNs. P$_{\rm{KS}}$ for the mass-matched control sample compared vary from
36\%\ in the low bin to 85\%\ in the intermediate bin, much higher than for the general control sample.

Simply matching in stellar mass leads to a distribution of U-B colors among inactive
galaxies that is much more similar to that of the AGN hosts. This is because X-ray detected AGNs are much more 
likely to be found in massive galaxies than low mass galaxies \citep[e.g.,][]{silverman09}. These massive hosts tend to be redder and have lower 
levels of star formation than the general population of field galaxies. 
Simply choosing a control sample to have the same range
of mass as the AGNs, as was done for the general control sample,
is not sufficient to account for stellar mass selection effects. 
Careful matching of the stellar mass distribution is essential.

We may conclude from this exercise that stellar mass selection effects are the primary 
cause of any differences in the color distributions of active and inactive galaxies. 
This is consistent with the results from the study of \citet{xue10} using the CDF-S 2MSec survey.

\subsection{AGN Hosts vs. Galaxies of similar \mass}

We now turn to a closer examination of the U-B colors of AGN hosts and mass-matched inactive galaxies.
The histograms in the right panels of Fig.\ref{colors_masses} show significant similarities, but 
some systematic differences remain. 
However, stochastic effects from the small size of the AGN sample and the large uncertainties in the 
colors of some galaxies can complicate our interpretation of these differences. 
In order to account for these effects, we `regularize' the \dcol\ distributions
of the AGNs and mass-matched control sample using the monte-carlo bootstrapping procedure (\S5.2). 
The results are shown in Fig.~\ref{color_trends}. The left panels of the figure 
plots the regularized distributions of \dcol\ for AGNs (solid histograms) 
and mass-matched inactive galaxies (open histograms) in the same three redshift bins. 
Error bars on the open histograms give the $1\sigma$ uncertainty on the distribution of the control sample. As
discussed in \S5.2, these errors take into account the stochasticity of the small AGN sample size as well as
modeling errors. 

We concentrate first on a comparison of AGN host color distributions across redshift. In the low bin, one finds a fairly
tight range in \dcol, spanning about 0.6 mag and peaking sharply at the Green Valley (\dcol$=0$). In the intermediate 
and high bins, the peak offset still remains roughly fixed. However, the 
width of the distribution changes considerably between the
higher redshift bins and the low bin, with a larger scatter towards red and blue colors. The largest increase
is in the fraction of galaxies with blue U-B colors (negative \dcol). 
Since we have removed cases where blue nuclear light could potentially contaminate
extended light in the AGN hosts, this evolution in color is likely a consequence of younger light-weighted mean 
ages in high redshift AGN compared to low redshift AGN, as well as evolution in the 
mean metallicity and dust obscuration with redshift. A very similar evolution in the U-B color distribution is also
evident in the mass-matched control sample. This underlines the fact that this
color evolution is not due to any increasing contribution from AGN light, as this would not be manifested 
among inactive galaxies. 
 
Comparing now the color distributions of AGN hosts and control galaxies
in the left panels of Fig.~\ref{color_trends}, we see that, by and large, the histograms are 
quite similar in the all the redshift bins, as was the case for the unregularized distributions. 
There may remain a marginal tendency for the modal colors of the AGNs to be bluer than the control sample, but
the differences in the overall distributions are minor and not significant, consistent with the high
P$_{\rm{KS}}$ values from the the K-S tests. 

While the general consistency between the colors of AGNs and inactive galaxies of the same stellar mass implies
that their stellar populations are also quite similar, certainly when compared to the general field population,
one may wonder if other effects could influence this interpretation. For example,
could differences in the level of dust obscuration in AGN hosts hide possible differences in stellar populations? 
For example, \citet{cardamone10} find that the application of a dust obscuration correction to the colors of AGN hosts 
at $z\approx1$ leads to a pronounced bimodality, paralleling the properties of normal Green Valley galaxies. 
To explore possible effects from dust obscuration, we plot, in the right panels of Fig.~\ref{color_trends}, 
distributions of the intrinsic \dcol\ for the AGN hosts and mass-matched control galaxies.
The intrinsic U-B colors come from the population synthesis fits after disregarding the fitted extinction. 

Comparing intrinsic and measured color distributions (right and left panels respectively), 
one sees that dust obscuration acts to
move some galaxies with very blue intrinsic colors (young mean ages) towards the red, 
tightening the overall color distribution of galaxies. In the right panels,
the color distributions of both AGNs and inactive galaxies are now much broader, with a much more pronounced
and bluer cloud of star-forming galaxies, especially among the inactive sample. 
However, the trends that appear in the left panels are also repeated in the right panels.
The significant tendency for AGN colors in the low bin to peak in the Green Valley remains, 
at odds with the results of \citet{cardamone10}. In the intermediate and high bins, the AGNs and
control galaxies show rather similar distributions, with a small, barely significant tendency to peak at
colors that are $\sim 0.1$ mag bluer than the peak of the mass-matched control galaxies, placing them
marginally closer to the Green Valley.
 
Before embarking on an interpretation of these color differences, we explore the degree to which such differences
arise simply from the limitations of the mass-matching process. The \cmd\ in the top panels of Fig.~\ref{extended_cmds}
indicates that fraction of blue galaxies increases greatly among the field galaxy population towards higher redshifts. 
Since we have allowed the mass-matched control sample to include galaxies that can be as much as a 
factor of two less massive than the AGN hosts, a larger fraction of lower-mass 
blue galaxies will make it into the control set towards progressively higher redshifts, 
due to greater errors in the stellar masses and the much greater density of blue galaxies. 
For these reasons, a general trend towards increasing numbers of blue galaxies among the inactive sample
compared to AGNs may be expected from the finite tolerances of the mass-matching procedure. 
\begin{figure}[t]
\figurenum{10}
\label{zerror}
\centering
\includegraphics[width=\columnwidth]{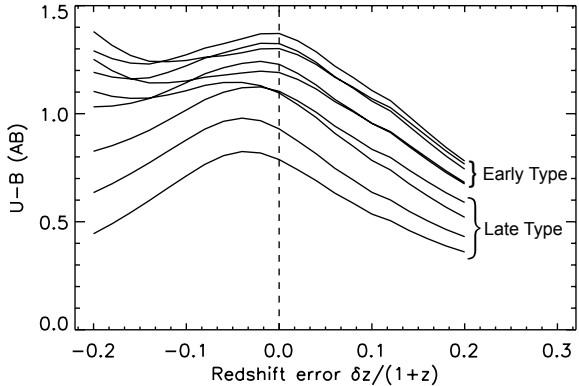}
\caption[Effect of zerrors]
{  The effects of redshift errors on the U-B colors of galaxies. Each line tracks the color of a certain spectral template
of a model galaxy as the redshift error $\delta z$ is increased or decreased by upto 20\% of $1+z$. 
Due to the sharpness of the Balmer/$4000$\AA\ break, redshift errors tend to lead to bluer colors, irrespective
of the sign of the error. The characteristic redshift error of galaxies in this study is $|\delta z/(1+z)| = 0.07$.
}
\end{figure}

\subsection{Effect of Redshift Errors}

The U-B color is sensitive to the strength of the 4000 \AA\ break, and hence to the light-weighted
mean age of the stellar population of a galaxy. However, in real galaxy catalogs with redshift errors, 
the error in the color can depend in a complex manner on the redshift errors of the galaxy. 
If the redshift of a galaxy is in error, 
the synthetic photometry from SED fits will span wavelengths
blue-ward or red-ward of the break, where the slope of the galaxy spectrum tends to be shallower
than across the actual break. Hence, redshift errors tend to lead to bluer U-B colors
irrespective of the sign of the error. 

This is potentially important since a large number of our high redshift AGN and more than 50\% of the control
galaxy sample do not have highly accurate spectroscopic redshifts, but instead have 
photometric redshifts with considerable uncertainty. The characteristic absolute value of $\delta z/(1+z)$ for galaxies
with photometric redshifts in the mass-matched control sample is 0.07, and around 
90\% have an estimated $|\delta z/(1+z)| < 0.1$
Hence, one may expect that model U-B colors for the objects with photometric redshifts would be 
preferentially bluer, which may affect the comparative color studies of previous sections.

\begin{figure}[t]
\figurenum{11}
\label{control_colors}
\centering
\includegraphics[width=\columnwidth]{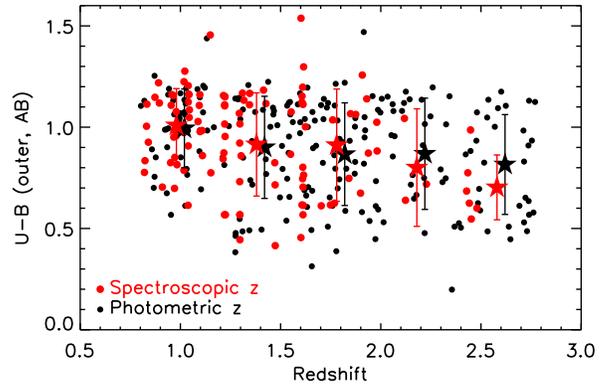}
\caption[Control sample colors with redshift]
{  Rest-frame outer U-B color vs. redshift of inactive galaxies from the mass-matched control sample. The black
points are galaxies with purely photometric redshifts, while the red points are galaxies with 
spectroscopic redshifts. The red and black points have very similar color distributions,
which implies that the effect of photometric redshift errors on the color distributions of the control sample
(see Fig.~\ref{color_trends}) are small to negligible. The mean color in five redshift bins are shown as large 
star-shaped points. Beyond a redshift of 2, the objects with spectroscopic redshifts tend to have slightly
bluer colors than objects with photometric redshifts, presumably because of selection effects that favor
blue star-forming galaxies with emission lines or strong Lyman breaks in such high redshift spectroscopic
samples. 
}
\end{figure}

The degree of this effect is shown in Fig.\ref{zerror}, in which we have plotted simulated U-B colors for a set
of representative galaxy spectral templates from the SWIRE template library \citep{polletta07}; three quiescent
galaxy templates with ages between 2-13 Gyr (`Ellipticals') and six star-forming galaxies with varying degrees of
current star-formation (`Spirals'). The galaxy templates were shifted in wavelength 
to simulate the effects of a redshift error $\delta z$, parametrized in the figure by the $\delta z/(1+z)$. The simulated
colors are reddest at $\delta z$ between -0.02 and 0.0, depending on the template, but get progressively
bluer as the redshift errors get both larger or smaller. Due to the sharpness of the $4000$\AA\ break, redshift errors 
tend to lead to bluer colors, irrespective of the sign of the error.

We constrain the importance of this effect on our rest-frame colors by comparing the U-B distributions of the 
mass-matched control galaxies with and without spectroscopic redshifts. In Fig.~\ref{control_colors}, we plot
the U-B colors of these two sets of control galaxies against redshift. The distribution of the
black points (galaxies with photometric redshifts) is very similar to those of the red points
(galaxies with spectroscopic redshifts) and clearly do not appear to be systematically bluer.
At the highest redshifts ($z \sim 2.5$), the objects with spectroscopic redshifts tend to be a little bluer,
probably because bright FUV continua or strong line-emission, both associated with blue
star-forming galaxies, are a necessary prerequisite for these spectroscopic redshifts.

From this we can conclude that photometric redshift errors for the control sample do not introduce strong systematics
in the U-B color distributions of the galaxies. A likely cause for the weakening of this effect
is that galaxies with prominent spectral features, such as breaks, usually have better determined photometric
redshifts as a consequence of the sharpness of the break, which works against the severity of the effect. 
The trends in Fig.~\ref{color_trends} are likely to reflect the true colors of the AGN and control galaxies.

\begin{figure*}[ht]
\figurenum{12}
\label{color_gradients}
\centering
\includegraphics[width=\textwidth]{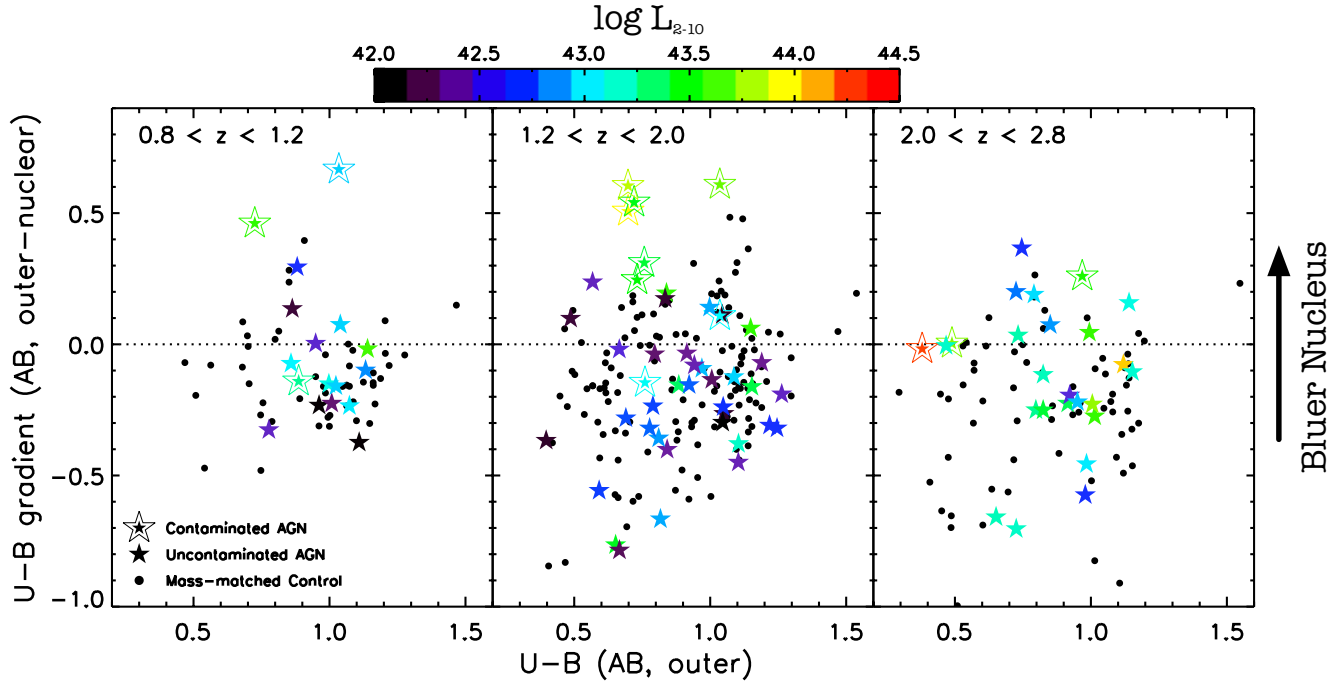}
\caption[Gradients vs. Outer colors]
{ Rest-frame U-B color gradients (outer - nuclear) plotted
against the outer U-B color, in three redshift bins. AGNs are shown by colored star points,
where the color represents the hard band X-ray luminosity \lhard\ (as shown by the color bar). 
Sources with nuclear contamination are marked with concentric star points. 
The small black points are the mass-matched control sample of inactive galaxies. 
}
\end{figure*}

\section{Color Gradients}


Radial color gradients are an important constraint on the role of nuclear activity in the regulation of 
star-formation in AGN hosts. If AGN are directly responsible for the shut down of star-formation
in massive galaxies, a difference is expected in the gradients of AGNs and inactive galaxies 
of a similar stellar mass. The time-scale over which feedback from an AGN blows out
a galaxy's gas will determine the steepness of the gradient. The luminosity of the AGN is
likely to regulate the scale over which the gradient may be measured - low luminosity AGN only
supply enough energy and radiation pressure feedback to influence the very inner parts
of a galaxy. 


In Fig.~\ref{color_gradients}, we compare the gradients in the rest-frame U-B colors of the AGNs 
(large colored star points) and mass-matched control sample (small black points) against their outer U-B colors, in 
two redshift bins. Gradients are calculated as the difference between the outer and
nuclear U-B colors, i.e, colors from fixed apertures of radius $0\farcs4$-$1\farcs0$ (annular) and $0\farcs1$ respectively. 
A lower/higher value of the gradient implies a bluer/redder outer color, with the dotted line showing a 
flat gradient. In addition, we use colors for the star points to indicate the X-ray luminosity of the AGNs, 
from L$_{{2\textrm{-}10}} = 10^{42}$ (black stars) to L$_{{2\textrm{-}10}} = 10^{44.5}$ (red stars). 
The AGNs with nuclear contamination are included in the figure as large concentric star points. 

A first observation is that the median gradient of both uncontaminated AGNs and control galaxies is
negative, with most points lying below the flat-gradient dotted line. Therefore,
both AGNs and inactive galaxies are generally redder in the central few kpc, compared to their outskirts.
The median offset from a flat gradient varies from $\approx -0.16$ in Bin A to $\approx -0.25$ in Bin C. The
scatter in the gradient also increases considerably towards high redshifts. The source of this change in
the median and the scatter could be intrinsic to the galaxy population, but it is also influenced by the
accuracy of PSF matching (to which the nuclear colors are particularly sensitive) as well as the larger
errors in photometry suffered by the more distant galaxies. This is why a careful control sample processed through
the same selection criterion, photometric measurement and modeling method is critical to constrain true differences
in the color gradients between AGNs and normal galaxies.

In general, the form and scatter of the distribution of color gradients between uncontaminated 
AGNs and inactive galaxies are similar, implying that the presence of the AGN does not strongly
influence the distribution of SF in their hosts. This, coupled with observation that
the outer colors of AGN hosts are consistent with those of normal 
galaxies of the same mass, suggests that the outer stellar populations of AGN hosts are 
the suppression of star-formation by feedback effects does not extend beyond
a few kpc, at least for AGNs in the luminosity range probed by our sample (L$_{{2\textrm{-}10}} \lesssim 45$ \ergs).

\section{Star-Formation Histories of AGN Hosts}

The detailed sampling of the SEDs of AGNs and control galaxies over 6-7 broad photometric bands 
enables a determination of the light-weighted star-formation histories (SFHs)
parametrized by a suite of increasing and decreasing $\tau$ models and constant star-formation rate models (\S5). 



Before proceeding with an examination of SFH parameters, we highlight the 
dominance of exponentially decreasing SFHs among AGNs and similar mass galaxies.
This is consistent with their high stellar masses and relatively red colors, even out to $z\sim3$. 
Among AGNs and control galaxies in the low bin ($0.8<z<1.2$), essentially none have
flat or rising SFHs for the best-fit models. The fraction of galaxies with non-declining SFHs increases in the
intermediate and high bins (13\% at $1.2<z<2.0$ and 23\% at $2.0<z<2.8$), but is still quite low.
The rates of best-fit rising SFH models are as common among X-ray AGN as among inactive galaxies.

The light-weighted stellar age is one of the fundamental physical characteristics of the SFH of galaxies.
However, typically the age is quite hard to estimate accurately as it can be degenerate with both the metallicity
of the stellar population and the level of dust extinction. Therefore, the confidence intervals on the stellar age
from our SED fits can be quite large and are typically skewed towards young ages. In order to make statistically
valuable conclusions about the age distributions of the AGNs, and their comparisons with normal galaxies, we
apply our bootstrap regularization procedure to arrive at more representative age distributions for the galaxies.  

\begin{figure}[t]
\figurenum{13}
\label{ages}
\centering
\includegraphics[width=3.5in]{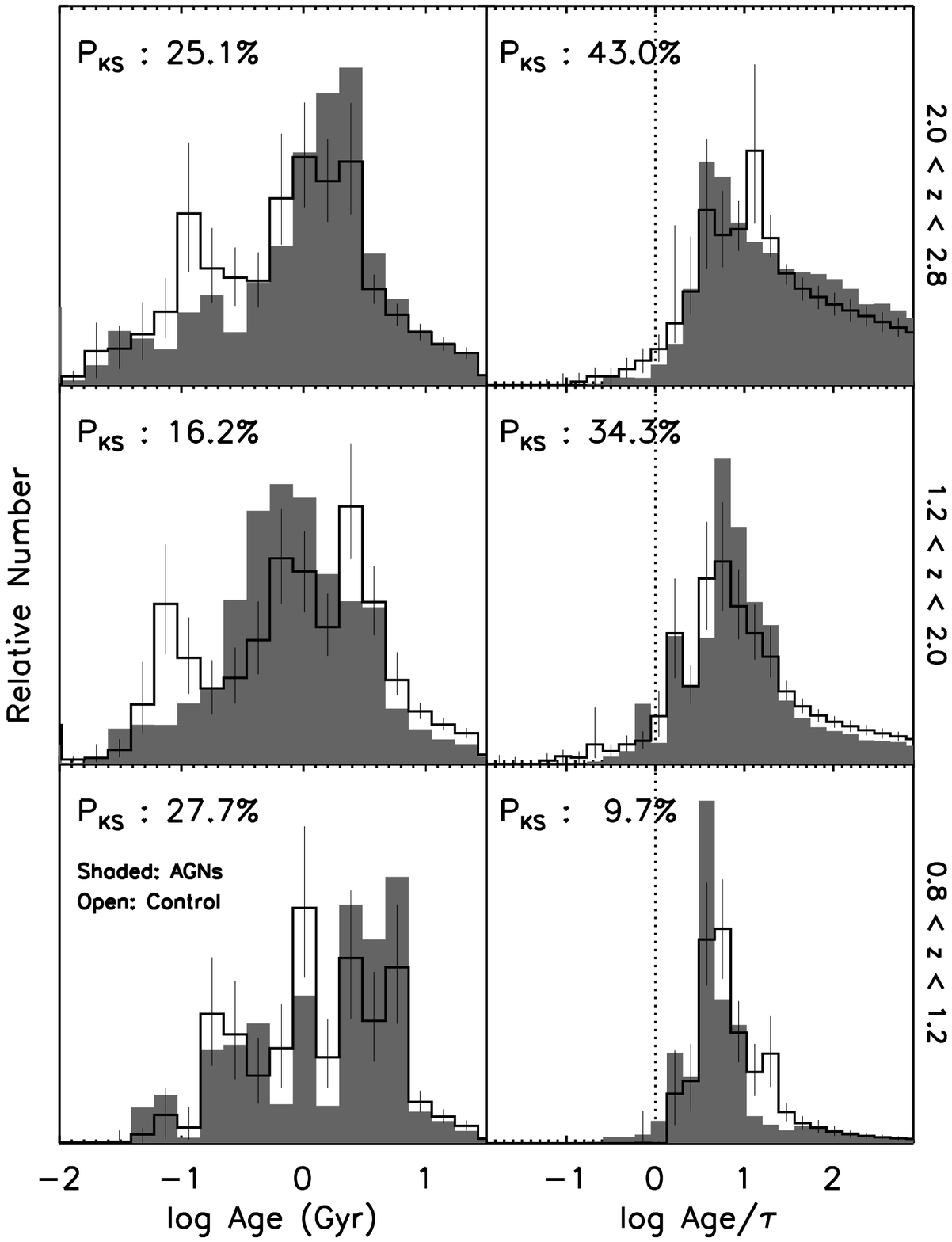}
\caption[Ages Distributions]
{ A comparison of the distributions of the light-weighted ages (left panels) and `normalized ages' (right panels)
of AGNs and control galaxies estimated from photometry in extended light apertures.
AGN hosts (grey histograms) are compared to mass-matched inactive galaxies (open histograms) in three
redshift bins, as indicated. Distributions have been regularized by a monte-carlo bootstrap procedure (\S5.2).
Error bars on the control galaxy histograms show the approximately $1\sigma$
variation in the distributions for sample sizes comparable to those of the AGN hosts. In each panel,
the two-sided Kolmogorov-Smirnov probability P$_{\rm{KS}}$, measured on the unregularized data,
compares the similarity of the two distributions.}
\end{figure}

The results are shown in the left panels of Fig.~\ref{ages}. As before, the AGNs are shown as shaded histogram, 
while the control sample are shown as an open histogram, in the low, intermediate and high redshift bins. 
The P$_{\rm{KS}}$ values in these panels come directly from K-S tests on the original distributions, not the regularized
distributions shown in the panels. They are a measure of the consistency of the age distributions without taking
into account the skewed uncertainties on best-fit ages.
In the low bin, the control galaxies show a relatively broad range of ages. The AGNs
have a distribution that is a bit narrower than the control sample, but not significantly. To the uncertainties
of the sample, the distributions are very similar.

In the higher redshift bins, the sample sizes of both AGNs (and, by extension, mass-matched control galaxies) 
are 2 to 3 times larger, making for better defined age histograms. 
The control galaxies in the intermediate and high bins show a broad but definite 
peak at 2 Gyr and another smaller narrow peak at $\approx 100$ Myr. Compared to the low bin, there is a larger
fraction of inactive galaxies with younger ages, stretching down to our lower age limit of 10 Myr. This is
consistent with the larger fraction of blue galaxies at higher redshifts among the mass-matched sample as well
as the AGN hosts.

The AGNs in the intermediate bin and high bins again appear to be quite similar to the control sample, though
in both bins, the AGN distribution is a bit narrower than the control. In addition, in both bins, the AGNs do not
show the second peak in the age distribution at young ages which is seen in the control sample. This
difference is fairly significant ($\approx 2\sigma$ at ages $\lesssim0.1$ Gyr), but may be 
related to the limitations of mass-matching at higher redshifts. 


We also explore, from our SFHs, the degree to which AGN hosts have built up their stellar content
at all these redshifts. We define a parameter called the 'normalized age', which is the ratio of stellar age to $\tau$.
This parameter, for declining $\tau$ models, indicates how much older a particular galaxy is compared to the timescale
over which most of its stars form. In the right panels of Fig.~\ref{ages}, we 
compare normalized age histograms of AGNs and control
galaxies in our three redshift bins, derived from our SED fits and regularized using the bootstrap procedure. 
Only objects with declining $\tau$ models are used in these histograms, since the
normalized age has a different interpretation for exponentially increasing or constant SFH models.

One may see that at all redshifts considered, the normalized age distributions for AGNs and control galaxies
peak at values greater than 1. The distribution is quite narrow at $z\sim1$, but get progressively broader 
at higher redshifts as a strong tail develops to high values. This is because $\tau$
is weakly constrained among faint high-redshift galaxies and low $\tau$ values lead to 
progressively higher normalized ages. 

In the intermediate and high bins, the peaks and widths of the AGN and control distributions are very 
similar.
In the low redshift bin, however, a K-S test suggests that the best-fit normalized ages of AGNs and control 
galaxies differ significantly, with a P$_{\rm{KS}} \lesssim 10\%$ that they are drawn from the same parent distribution.
The regularized histograms at normalized ages $\gtrsim 10$ differ at the level of a few $\sigma$, showing that 
the AGNs display a narrow distribution of normalized ages and a weaker tail to long ages than the control
sample. In essence, this means that at $z\approx1$, the normalized ages of AGNs are slightly 
shorter and have less variation than inactive galaxies of the same mass.

\section{Discussion}

\subsection{The Colors and SFHs of AGN Hosts}

WIth the high spatial resolution and unparalleled sensitivity of the WFC3 camera, we have examined the 
photometric properties and star-formation histories of AGN host galaxies in the CANDELS/CDF-S field at 
redshifts of $0.8<z<2.8$, while avoiding complications arising from strong nuclear contamination.
Consistent with earlier studies, X-ray selected AGNs are found to reside in galaxies that are significantly more
massive than the general field galaxy population \citep[e.g.,][]{silverman09}. 
The typical stellar mass of AGN hosts is log M$_{*} \sim 10.5$
and it changes very little (less than a factor of 3) towards high redshifts. This suggests 
that low- to intermediate-luminosity AGNs reside in hosts of a characteristic stellar mass over a large
range in redshift. 

In keeping with their high stellar masses, AGN hosts are more likely to be found 
in galaxies with red or intermediate colors, compared to the general population of field galaxies.
To account for covariances in galaxy properties that are a strong function of stellar mass, we
make a detailed comparison of the colors, color gradients and SFHs of AGN hosts to a
comparison sample of inactive galaxies that are matched in stellar mass to the AGNs. 

We find that, to a major degree, AGN hosts at all redshifts have basic photometric properties -- 
rest-frame colors, extinctions and color gradients -- that are similar to mass-matched inactive galaxies.  
AGN hosts get considerably bluer and more compact towards higher redshifts, but this evolution is 
consistent with the changes seen in the inactive galaxy population as well. In addition, both AGNs and 
inactive galaxies show typically old stellar populations (ages $\gtrsim$ 1 Gyr) and are better fit by models 
with a declining SFH at all redshifts (though the fraction of galaxies with flat or rising SFHs increases with redshift). 
The characteristic age of AGNs is long compared to the timescale of SF (i.e, $\tau$ for exponentially declining SF models). 
In essence, low- to moderate-luminosity AGNs are found in relatively normal massive galaxies at all redshifts to $z=3$.

\subsection{AGNs and the Transformation of Galaxies}

Several studies have suggested that the high frequency of AGN in the Green Valley is evidence of a close association between the quenching of star formation in galaxies and the feedback from active nuclei \citep{nandra07, schawinski07,
schawinski10}. According to this scenario, the energy output of the accreting SMBH couples with the gas in the disk of a star-forming blue cloud galaxy and either drives most of it out of the galactic potential \citep{kaviraj11}, or, alternatively, heats it to a temperature beyond $10^6$ K. At this point, the gas is too hot or too diffuse to cool back into the galaxy disk and the star-formation shuts down. The time-scale of AGN-driven quenching is short enough (~0.5 Gyr) that the galaxy's light is dominated by an intermediate age stellar population with colors typical of the Green Valley. The relative infrequency of Green Valley galaxies, reflected in the color-bimodality of the galaxy population, is a consequence of the rapidity of AGN-driven quenching. A similarity between the estimated time-scales of AGN activity and the time-scales needed to move a quenched galaxy onto the Red Sequence seems to imply a close association between the two processes \citep{ schawinski07,bundy08}. \citep{ schawinski07} also find close associations between a population of local blue early-type galaxies and AGN activity, indicating that AGN may play a role in the quenching of ongoing star-formation in such systems.

However, an elevated fraction of AGNs in the Green Valley can also arise from a simple phenomenological explanation, which does not require a strong role for AGN feedback in the transformation of galaxies \citep{sanchez04, silverman09}. Two key ingredients necessary for producing an AGN are a) fuel, in the form of relatively cold inter-stellar gas in the host galaxy, and b) a supermassive black hole. For an accreting black hole of a given mass to be detected at intermediate-to-high redshifts in a typical deep flux-limited X-ray survey, it either needs to have a high accretion rate (or Eddington ratio) or a high black hole mass, or both. Since massive black holes are typically found in galaxies with massive spheroids \citep[e.g.,][]{haring04}, the detection rate of X-ray AGN will increase with bulge fraction and stellar mass, but to a point. The most massive galaxies at essentially all redshifts are almost always elliptical galaxies, which have hot haloes and are deficient in cold gas, due to heating in virial shocks \citep{birnboim07}, feedback from radio jets \citep{best06, croton06} or strong `quasar mode' feedback during a possible earlier major merger episode \citep[e.g.,][]{urrutia08}. Such systems are unlikely to host black holes that accrete significant quantities of cold gas. Therefore, in this picture, X-ray AGNs are preferentially found in galaxies with massive spheroids (and black holes), but also with disks which contain sufficient cold gas to keep up regular cycles of accretion. Such galaxies are typical of the Green Valley \citep{cassata07}. In this scenario, the prevalence of AGNs in galaxies with intermediate colors is solely a consequence of the necessary conditions for an accreting black hole to be detected in X-rays, coupled with a static or redshift-dependent relationship between SMBHs and spheroids \citep[e.g.,][]{tremaine02, merloni10, bennert11, kormendy11}, which is probably set during high accretion rate phases, such as through galaxy mergers \citep{hopkins08a}. Considerable scatter in the location of AGNs in the color-mass diagram would be naturally expected, mostly towards redder galaxy colors, since some early-type galaxies are known to contain sizable amount of cold dusty nuclear gas \citep{ferrarese06}, possibly through settling of dense gas in hot atmospheres or mass loss from evolved stars. 

If AGN hosts are indistinguishable from a population of massive inactive galaxies of the same stellar mass, this would support the latter scenario. In this scenario, the characteristics of AGN hosts would be set not by their nuclear activity, but by the strong dependence of galaxy properties on stellar mass, modulated by the conditions necessary to fuel SMBH accretion (massive bulge and gas-rich disk). On the other hand, if AGN hosts deviate from inactive galaxies of the same stellar mass in a systematic way, this could indicate that nuclear activity does in fact play a role in determining the gross properties of the hosts and lend support for a more intimate two-way connection between AGNs and their hosts.

Unfortunately, our study is somewhat inconclusive in this regard. We can say for certain that a large part of 
the typical properties of  AGN hosts are set by their stellar mass, rather than the fact that they host an 
accreting SMBH. This is clear since AGNs for the most part are really quite similar to inactive galaxies of 
the same mass. This rules out any models that postulate strong differences between the AGN and 
normal galaxy population, for e.g., models that tie most nuclear activity to galaxy mergers or post-mergers.
Such a result is consistent with recent studies that find that AGNs have very similar morphologies to 
inactive, mass-matched galaxies across redshifts we consider in this work \citep{cisternas11, schawinski11, kocevski12}. 

We notice some possible minor differences at $z\sim1$ -- for e.g., the tendency for
AGN hosts to have a narrow range in U-B color and have shorter normalized ages than inactive galaxies -- 
which should not be strongly affected by systematics in the construction of a control sample. These results
are qualitatively consistent with earlier studies of AGN hosts at $z\sim1$ \citep[e.g.,][]{sanchez04, gabor09},
though stellar mass selection effects were not fully taken into account in much prior work. In general,
AGNs have a tendency towards older stellar ages, but a slightly higher fraction of them 
have formed stars more recently. These two apparently opposing notions 
may be reconciled if AGN hosts have older formation times but broader SF timescales than inactive 
galaxies. In other words, they have formed a small fraction of stars more recently, possibly in association 
with the process that drives accretion.

At higher redshifts, these differences become less pronounced and more susceptible to systematics. 
Our intermediate redshift bin ($1.2<z<2.0$) has the best statistics, low incompleteness 
and stellar mass uncertainties that are not too severe. In this bin, we find that AGNs are most 
closely similar to inactive galaxies. 

If AGNs are no different from inactive galaxies, or associated with low levels of recent star-formation, 
then the importance of AGNs as mechanisms for driving the prompt transformation of galaxies 
from star-forming to quiescence may be overstated. Such models predict that AGNs are in galaxies with 
less recent SF, not more, since the phase of AGN activity leads to the immediate quenching of
star-formation, while in galaxies without AGNs, star-formation can proceed unabated. 
However, a possible variation on this scenario may be able to reconcile our observations with a role for 
AGN feedback. If AGN activity is synchronized with star-formation phases, 
it is possible that feedback may effectively shorten the timescale of star-formation and therefore reduce the efficiency of 
star-formation in these massive galaxies. In such a scenario, AGN hosts contain signatures of recent 
star-formation because it is only in galaxies with such recent star-formation 
that AGNs have been triggered, possibly by the same process that triggered the formation of stars as well.
Such processes could be, for e.g., minor mergers or satellite interactions, 
bar instabilities or the infall of fresh gas from the inter-galactic medium. The feedback 
from the active nucleus would then cut short a burst of star-formation that would have 
continued longer had the AGN not been present. In this way, by gradually eroding 
the duty cycle of star-formation in massive galaxies, AGNs may be able to play a role in their transformation. 

\subsection{Evolution in the Relationship of AGN Hosts to Normal Galaxies?}

From a large study of the morphologies and colors of AGN hosts in the Galaxy Zoo survey, \citet{schawinski10}
find that local AGNs satisfy a strong preference for the Green Valley, even when mass selection effects are
taken into account. In particular, the most massive local galaxies are less likely to host AGNs compared
to galaxies that have bluer colors and lower masses. A recent study of the properties of a hard X-ray selected
BAT sample of AGNs \citep{koss11}, which span a luminosity range that is similar to the CDF-S 4 MSec sample, 
also find a similar result, while again accounting for stellar mass selection effects. Taken at face value,
these studies suggest that local AGNs are still preferentially found in the Green Valley and may be more
closely associated with a quenching galaxy population.

This contrasts considerably with what we find at $z>1.5$, where AGN hosts 
have a similar spread in color and SFH as inactive galaxies of the same mass,
and indeed are very common among the most massive systems at these redshifts. 
At $z\sim1$, our results suggest a situation that is intermediate between these local studies and the
properties of AGN hosts at higher redshift. If true, we may be witnessing a change with redshift
of the relevant processes that relate substantial black hole accretion to the properties of the
host galaxy. The high gas fractions and turbulent, clumpy gas disks seen among massive
galaxies at $z\sim2$ should allow more frequent and consistent inflow of gas to the central SMBH \citep{bournaud11}.
Such processes will be able to occasionally fuel even more luminous phases, such as quasars, which
require disruptive processes like major mergers to drive them at low redshifts. 
If so, the relative importance of major mergers as the triggers of major SMBH growth at high redshift will decline, 
while secular processes, such as turbulent accretion, will be more important. 
This may explain the change in the association between galaxies and AGN activity towards higher redshift:
if secular inflow can fuel most AGNs at $z\sim2$, all galaxies with such gas disks will be able to
accrete significant quantities to be able to shine as an X-ray AGN. In this case, every massive
galaxy is a candidate AGN host. At lower redshifts, only massive spiral galaxies, with enough
gas in their disks, can fuel secular AGN phases, while a larger fraction of Seyfert activity is fueled
in post-merger systems \citep{schawinski10}. This will naturally lead to an evolution in
the make-up of AGN hosts towards higher redshifts. 

A deeper understanding of the physics of feedback, from simulations \citep[e.g,][]{debuhr10} 
and observations of active galaxies across a range of scales, is crucial towards constraining 
whether, how, where and when AGNs play a part in altering the nature of their hosts. Our 
study allows us to rule out the importance of strong and prompt quenching in AGNs but weaker, 
longer lasting and more pervasive processes could still have an critical role in modulating the transformation
of galaxies. In addition, we suggest that evolution in the principal mode of AGN fueling may be relevant
in studies of AGN hosts with redshifts. As the CANDELS survey expands, we will be able to study
$\sim$1000s of AGNs across a wide range of redshifts with this remarkable dataset and, hopefully,
tease apart the various roles of galaxy mass, gas inflow, AGN luminosity and feedback in the evolution
of active galaxies.

\acknowledgements

We thank S. M. Ammons and F. Bournaud for fruitful discussion. 
This work is based on observations taken by the CANDELS Multi-Cycle Treasury Program with the NASA/ESA HST, 
which is operated by the Association of Universities for Research in Astronomy, Inc., 
under NASA contract NAS5-26555. This research has made use of data obtained from the 
Chandra Data Archive and the Chandra Source Catalog, and software provided by the 
Chandra X-ray Center (CXC) in the application packages CIAO, ChIPS, and Sherpa.

\bibliographystyle{apj}
\bibliography{candels_xray}

\appendix
\section{Effects of AGN Point Source Contamination on Aperture Photometry}

The CANDELS WFC3 F160W Point Spread Function has fairly bright wings relative to its core, particularly when compared
to the ACS PSF (95\% encircled energy radii of 0.8"/1.5" for ACS F814W/WFC3 160W). Therefore, one must understand the 
effects of scattered light on aperture photometry before interpreting such measurements, especially in 
AGNs, which can display strong nuclear point sources. 

We approach this in a two-part manner. We first predict the expected luminosity of AGN point sources in the rest-frame U-band and compare
these predictions to the luminosities of massive galaxies at the redshift range of our sample. From this input, we evaluate the range of
point source-to-galaxy flux ratios which our sample of AGNs are expected to show in the bluest bands. Next, we use a suite of 
simulated galaxies with added central point sources to evaluate the effects of AGN contamination
on extended aperture photometry as a function of galaxy size, sersic index and point source fraction.

\begin{figure*}[ht]
\figurenum{A}
\label{contam_tests}
\centering
\includegraphics[width=3.0in]{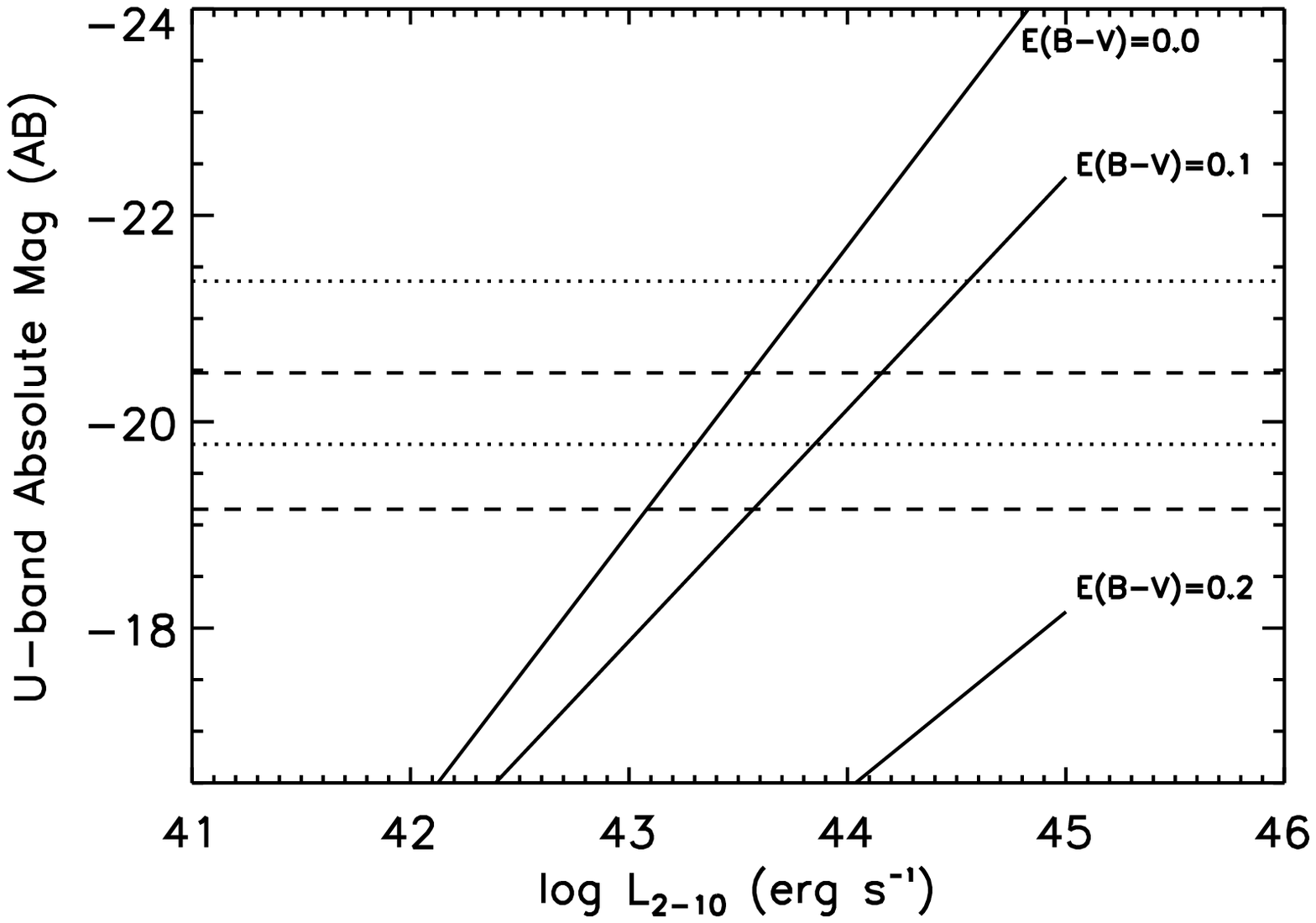}
\includegraphics[width=3.0in]{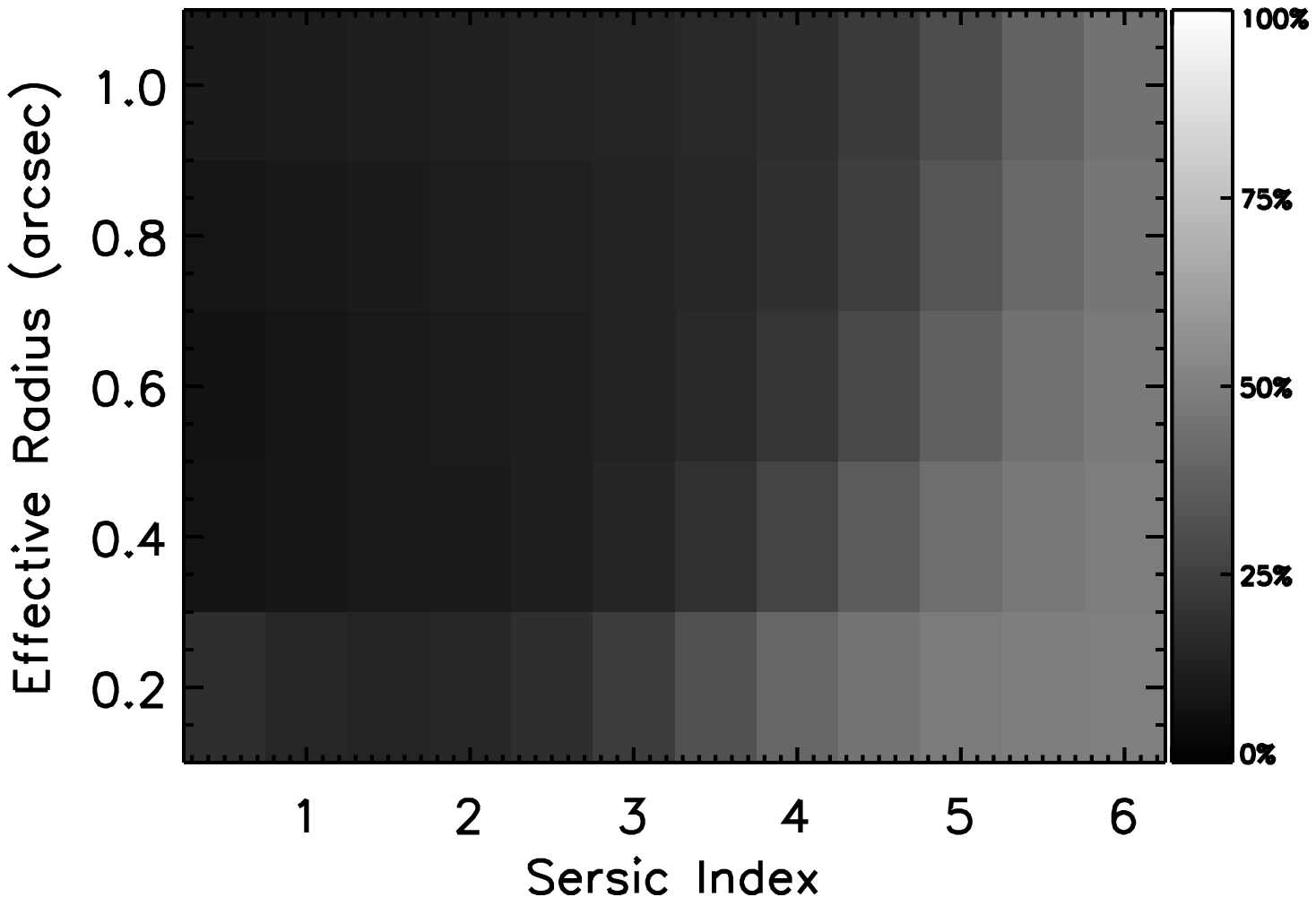}
\caption[Contamination tests]
{ Left panel: The U-band absolute magnitude of nuclear emission from an AGN as a function of the X-ray luminosity
of the AGN, for three values of nuclear dust extinction $E(B-V)$. These estimates were made assuming the X-ray to
mid-IR correlation of local AGN from \citet{gandhi09} and a mean Type I AGN SED to connect the UV to the mid-IR.
The typical range of U-band luminosities of normal massive galaxies in FIREWORKS at  $0.8<z<1.2$ (dashed lines)
and $2.0<z<2.8$ (dotted lines) are shown. At modest levels of extinction, even the most luminous AGNs in our
sample are too faint to adversely contaminate the photometry of the host. Only if an AGN is quite unobscured can
be dominate the galaxy's integrated photometry.

Right panel: The fractional increase in the annular flux over the light of a galaxy due to the presence
of a bright nuclear point source, for model galaxies with a range in size and S\'ersic profile index. The 
circular annulus spans galaxy radii from $0\farcs4$--$1"$, matching the annulus used for extended light
photometry in this paper. In the case
shown here, the point source is assumed to contain as much flux as the galaxy. Such AGN can be easily
identified visually unless the point source is reddened considerably by dust. For the luminosity range
of AGNs in our work, dust extinction will also strongly reduce the UV contribution any point sources. 
AGN contamination is therefore unlikely to be a large source of systematics in this study.
 }
 \end{figure*}

The hard-band X-ray luminosity of local AGNs is known to correlate quite well with their nuclear MIR luminosity over a large
range in intrinsic AGN luminosity. Specifically adopting the \lhard\ to 12.3 $\mu$m relationship from \citet{gandhi09},
and using model Type 1 AGN SEDs to connect the rest-frame U band luminosity to the MIR, we can predict
the relationship between \lhard\ and $M_{U}$ for the X-ray luminosity range of our AGN sample. The effects
of nuclear dust obscuration can also be explored by applying a range of extinction to the model AGN SEDs. We adopt a 
moderate luminosity Type 1 AGN SED from \citet{silva04} and Galactic screen extinction (the results are not very
sensitive to the choice of extinction law or mean AGN SED template).
The left panel of Fig.~\ref{contam_tests} summarizes this exercise. For AGNs with a given hard-band X-ray luminosity, we plot solid lines
which show the expected U-band luminosity for AGNs as a function of \lhard\ for three different values of E(B-V).
In addition, we show the 1$\sigma$ range in absolute U magnitude for inactive galaxies with stellar masses similar to the AGNs
in the redshift ranges $0.8<z<1.2$ and $2.0<z<2.8$ (dashed and dotted lines). 

From the figure, we deduce that \emph{unobscured} AGNs (E(B-V)=0) with \lhard$>10^{43}$ \ergs\ have point source 
luminosities that rival the total luminosity of galaxies across all redshifts in this work. However, even small levels of 
nuclear obscuration quickly depress the rest-frame U-band luminosity of the AGN, such that
point sources with a modest E(B-V)$=0.2$ are more than a magnitude fainter than most galaxies, even among the most
luminous AGNs in our sample ( \lhard$\sim10^{45}$ \ergs). Since the intrinsic emission from an AGN rises sharply
through blue and UV wavelengths, we expect the nuclear colors of AGNs with strong nuclear point source contamination
to be overwhelmingly blue. Obscured AGNs may display red point sources, by they are not expected to be very
luminous in comparison to the host galaxy for the nuclear luminosities and redshifts considered by this study.

This is supported by observations. X-ray selected AGNs
at $z<1.5$ in the CDF fields appear to show pervasive central point sources, attributed by some 
authors to highly reddened nuclear emission \citep{ballo07, simmons11} . The typical extinctions of these point sources were 
E(B-V)$>0.1$ and they accounted for lower than 10\%\ of the total luminosity of the host.
\citet{gabor09} performed a similar exercise on AGNs from the COSMOS field, which tend to
more luminous than the sample considered in this work. Even so, only a handful of AGN point sources from their study
were as luminous as their host galaxies.

For the second part of our analysis of AGN contamination, we simulated a suite of model circular galaxies
with Sers\'ic light profiles (index $n=0.5-6$) and half-light radii in the range \reff$= 0\farcs2-1\farcs5$. After
convolving the galaxies by the WFC3 F160W PSF (described in \S2.1), we added a point source to the
images, scaled to a total point-source to galaxy flux ratio in the range $R_{PS} = 0.01 - 2.0$.
In the right panel of Fig.~\ref{contam_tests}, we plot the fractional increase in flux in the extended light aperture 
($0\farcs4$--$1\farcs0$, used for all photometry in this paper)
due to a central point source of $R_{PS}=1$. 
Sources with such bright point sources would be easily identified visually as bright blue compact 
structures in the centers of these galaxies.

For galaxies with typical sizes (\reff$> 0\farcs4$) and sersic indices (1-4) found among
AGN hosts, a bright point source adds around a few 10s to 50\% more light in this aperture over the
light from the galaxy. Among AGNs with such strong nuclear emission,
our extended light photometry is contaminated by the central source less than the integrated
photometry of the galaxy. Since point sources are generally more compact than galaxies, 
extended light among fainter AGNs are proportionately contaminated even less than bright AGNs.
Through the application of our X-ray parameter-based cuts to exclude luminous, unobscured AGN,
we remove all cases of strong nuclear point sources. Among the remaining AGN, nuclear contamination
in the outer apertures is quite minimal and will not strongly influence our results.

\end{document}